\documentclass[journal]{IEEEtran}
\usepackage{dsfont}
\usepackage{graphicx}
\usepackage{subfigure}
\usepackage{amsfonts,amssymb}
\usepackage{lipsum}
\usepackage{cite}
\usepackage{bm}
\usepackage{subeqnarray}
\usepackage{algorithm}
\usepackage{algorithmic}
\usepackage{caption}
\usepackage{lipsum}
\usepackage{multicol}
\graphicspath{{figures/}}
\usepackage[cmex10]{amsmath}
\usepackage{amsfonts,amssymb}
\usepackage{epstopdf}
\usepackage{amsmath}
\usepackage{bm}
\usepackage{algorithmic}
\begin{document}
%

\title{Towards Optimal {Adaptive} Wireless Communications in Unknown Environments}

\author{\normalsize Pan Zhou, \emph{Member, IEEE} and Tao Jiang, \emph{Senior Member, IEEE}
\thanks{
This work was supported in part by National Science Foundation of China with Grants 61401169, 61531011, 61428104,  and 61325004, and Joint Specialized Research Fund for the Doctoral Program of Higher Education (SRFDP) and Research Grants Council Earmarked Research Grants (RGC ERG) with Grant 20130142140002.

Pan Zhou and Tao Jiang (corresponding author) are from the School
of Electronic Information and Communications, Huazhong University of
Science and Technology, Wuhan 430074, China (email: panzhou@hust.edu.cn;  tao.jiang@ieee.org).



%


} \\}

\maketitle
\thispagestyle{empty}
\pagestyle{plain}

\begin{abstract}
Designing efficient channel access schemes for wireless communications without \emph{any} prior
 knowledge about the nature of environments has been a very challenging issue, in which the channel state distribution
  of all spectrum resources could be entirely or partially stochastic or adversarial at different time and
  locations. In this paper, we propose an online learning algorithm for adaptive channel access of wireless communications in unknown environments based on
  the theory of multi-armed bandits (MAB) problems. By automatically tuning two control parameters, i.e., learning rate and exploration probability, our algorithms could find the optimal channel access strategies and achieve the almost optimal learning performance over
 time in different scenarios.  The quantitative performance studies indicate the superior throughput gain when compared with previous solutions and
  the flexibility of our algorithm in practice, which is resilient to both
  oblivious and adaptive jamming attacks with different intelligence and attacking strength that ranges from no-attack to the full-attack of
  all spectrum resources. We conduct extensive simulations to validate our theoretical analysis.
\end{abstract}


\begin{keywords}
Online learning, jamming attack, stochastic and adversarial bandits, wireless communications, security.
\end{keywords}

%
\IEEEpeerreviewmaketitle

\section{Introduction}
The design of channel access schemes is a pivotal problem in wireless communications.
Stimulated by  the recent appearance of smart wireless devices with adaptive and learning abilities, modern wireless communications have  raised very high requirements to its solutions, especially in complex environments, where accurate instant channel states can barely be acquired before transmission and long term channel evolution process are unknown (e.g., cognitive radio, smart vehicular  and military communications). Thus, it is critical
for wireless devices to learn and select  the best channels to access in general unknown wireless environments.

Many recent works have tackled the
channel access problem in unknown environments by online learning approaches, almost all of which are well formulated as the Multi-armed bandit (MAB) problem \cite{Bubeck12} due to its
inherent capability in keeping a good balance between ``exploitation" and ``exploration" for the selection of  channels and the superior throughput gain with the finite-time optimality guarantee, e.g., \cite{WangInfo11,ICNP11, XYinfocom11, XYTMC13, Yi11,XYICDCS14, YiToN12, ZhengInfocom13, MaghTVT2014, ZhengTSP14, QianJSAC12,ZhaoInfocom10}.  The main goal is
to  find a channel access strategy that achieves the optimal expected throughput by minimizing the term ``\emph{regret}" as learning
performance metric, i.e., the performance gap between the  proposed strategy and the optimal fixed one known in hindsight, accumulated over time.
 Briefly speaking,
 these works can be categorized into two different types of MAB models, namely, adversarial (non-stochastic) MAB \cite{WangInfo11,ICNP11,XYinfocom11}\cite{MaghTVT2014,QianJSAC12} and stochastic MAB \cite{XYTMC13,Yi11,YiToN12,XYICDCS14,ZhengInfocom13} \cite{ZhengTSP14,ZhaoInfocom10}. Stochastic MAB assumes that the channel state follows some unknown i.i.d. process, while adversarial MAB assumes that
the channel state can be controlled arbitrarily by adversaries (e.g., jamming attackers)
where its distribution is not i.i.d (i.e., non-i.i.d.) any more. Accordingly, the analytic approaches and performance results
 of the two models are distinctively different.
  A well-known truth is that  stochastic MAB  and  adversarial MAB have  regrets
  of order ``logarithmic-t"  \cite{Robbins1985} and ``root-t" \cite{non_MAB02} over time $t$, respectively. Obviously, the learning performance of the stochastic MAB
  highly outperforms that of the adversarial MAB.


 As we know, one key assumption of almost all existing works is the nature of environments, as a known prior,  is either stochastic or adversarial.
 This is limited in describing general wireless environments in practice, although it largely captures the main characteristic of them.  Because, in many practical wireless applications, the nature of the environment  is not restricted to either the stochastic or
the adversarial type, and it usually can not be known in advance.


On the one hand, the application of existing models may lead to bad learning performance when no prior
about the environment is available. Consider a wireless network deployed in a potentially  hostile environment. The number and locations of attackers are often unrevealed
to the wireless networks. In this scenario, most likely, certain
 portions of spatially-dispersed channels  may (or may not) suffer from denial of service attackers that are adversarial, while others  are stochastic distributed.  Compared with the classic mind that uses the adversarial MAB model \cite{WangInfo11,ICNP11,XYinfocom11}\cite{MaghTVT2014,QianJSAC12},
 here to design optimal channel access strategies, the use of the stochastic MAB  may not be feasible due to the existence of
 adversaries.  Meanwhile,  the use of the adversarial MAB model on all channels will lead to
 large values of regret, since a great portion of channels can still be stochastically distributed. Thus, it is hard to decide the type of MAB models to be used in the first place.

On the other hand,  the channel access strategy
based on stochastic MAB model \cite{XYTMC13,Yi11,YiToN12,XYICDCS14,ZhengInfocom13} \cite{ZhengTSP14,ZhaoInfocom10}  will face practical implementation issues, even though it is certain that there is no long term adversarial behavior. In almost all wireless communication systems, the commonly seen occasionally disturbing events would make the stochastic channel distributions contaminated. These include the burst  movements of individuals, the jitter effects of electronmagnetic waves, and the seldom but
 irregular replacement of obstacles, etc. In this case, the channel
 distribution  will not follow an i.i.d. process for a small portion of time during the whole  learning period. Thus, whether the  stochastic MAB theory is still applicable, how the contamination affects the learning performance
 and to what extend the contamination is negligible are not clear to us. Therefore, the design of a unified channel access scheme without any prior knowledge of the operating environment is very challenging. It is highly desirable and bears great theoretical value.

  In this paper, we propose a novel adaptive multi-channel access algorithm for wireless communications that achieves near-optimal learning performance without \emph{any} prior knowledge about the nature of the environment, which provides the first theoretical foundation of scheme design and performance characterization for this challenging issue. The proposed algorithm neither  needs to distinguish the stochastic and adversarial
MAB problems nor requires the time horizon for run.  To the best of our knowledge, ours is the first work  that bridges
the \emph{stochastic} and \emph{adversarial} MABs into a unified framework with promising applications in practical wireless systems.

The idea is based on the famous EXP3 \cite{non_MAB02}  algorithm in the non-stochastic MAB by introducing a new control
parameter into the exploration probability for each channel. By joint control of learning rate and exploration probability, the proposed algorithm achieves almost optimal learning performance in different regimes. When the
environment happens to be adversarial,  our proposed algorithm enjoys the same behavior as classic adversarial MABs-based algorithms and has the optimal regret``root-t" bound in the adversarial regime. When the environment happens to be stochastic, we indicate a problem-dependent
``polylogarithmic-t" regret bound, which is slightly worse than the optimal ``logarithmic" bound in \cite{Robbins1985}. Furthermore, we prove that
the proposed  algorithm retains the ``polylogarithmic-t" regret bound in the stochastic regime  as long as on average the contamination over all channels does not reduce the gap $\Delta$ between the optimal and suboptimal channels by more than a half. Note that all regret bounds are \emph{sublinear} to time horizon, which indicates the optimal channel access strategy is achievable. Our main contributions are summarized as follows.


\emph{1)}  We categorize the features of the general wireless communication environments mainly into four typical regimes: the adversarial regime,
the stochastic regime, the mixed adversarial and stochastic regime,  and the contaminated stochastic regime. We  provide solid theoretical results for them, each of which achieves the almost optimal regret bounds.

\emph{2)}    Our proposed AUFH-EXP3++  algorithm considers the statistical information sharing of a channel that belongs to
     different transmission strategies, which can be regarded as
    a special type of combinatorial semi-bandit\footnote{The term first appears in \cite{OR2014}, which means
 the reward of  each item within the combinatorial MAB strategy as a played arm will be revealed to the decision maker. } problem. In this scenario,
  given the size of all channels $n$ and the number of receiving channels $k_r$, AUFH-EXP3++ achieves the regret of order $O({k_r}\sqrt {tn\ln n} )$
  in the  adversarial regime (for usually considered \emph{oblivious} adversary) and the regret of order $\tilde{O}(\frac{{n{k_r}\log {{(t)}}}}{\Delta })$ in other stochastic regime up to time $t$. From the perspective of parameters
  $n$ and $k_r$ for different configurations of wireless communications, AUFH-EXP3++ achieves tight regret bound in
both the adversarial setting  \cite{OR2014} and the stochastic setting \cite{Branislav2015}. We also study  the performance of our
 algorithm under \emph{adaptive} adversary for the first time.

\emph{3)}  We provide a computational efficient enhanced version of the AUFH-EXP3++ algorithm.
Our algorithm enjoys linear time and space complexity in terms of $n$ and $k_r$ that indicates very good scalability, which can be implemented in large scale wireless communication networks.

\emph{4)}  We conduct plenty of diversified numerical experiments, and simulation results demonstrate that all  advantages of the AUFH-EXP3++  algorithm in our theoretical analysis is real and can be implemented easily in practice.


%




The rest of this paper is organized as follows: Related works
  are discussed in Section II. Section III describes the communication model,  problem formulation, and the four regimes. Section IV introduces  the optimal adaptive uncoordinated frequency hopping algorithm, AUFH-EXP3++. The performance results for different
  regimes are presented in Section V, while their theoretical proofs are shown in Section VI. Section VII presents a computational
  efficient implementation of the AUFH-EXP3++ algorithm. Numerical and simulation results are available in Section VIII. Finally, we conclude the paper in Section IX.

\section{Related Works}
Recently, online learning-based approaches to address wireless communications and networking problems in unknown
 environments have gained growing attention. The characteristics of learning by repeated interactions with environments are  usually categorized
 into the domain of reinforcement learning (RL). It is worth pointing out that there exists extensive literature in RL, which   generally
target  at a broader set of learning problems in Markov
Decision Processes (MDPs) \cite{Ref98}. As we know, such learning algorithms
 can guarantee optimally only asymptotically to infinity, which cannot be relied upon in mission-critical applications.   However, MAB problems
constitute a special class of MDPs, for which the regret
learning framework is generally viewed as more effective both
in terms of convergence and computational complexity.
 Thus,  the use of MAB models is
 highly identified. The works based on the stochastic MAB model often consider about the stochastically distributed channels in benign  environments, such as dynamic spectrum access \cite{Yi11}\cite{ZhaoTSP10}\cite{ZhaoInfocom10},  cognitive radio networks \cite{XYTMC13},  channel monitoring in infrastructure wireless networks \cite{ZhengInfocom13}\cite{ZhengTSP14}, wireless scheduling \cite{YiToN12}, and channel access scheduling in multi-hop wireless networks \cite{XYICDCS14}, etc.  The adversarial  MAB model is applied
 to adversarial channel conditions, such as the anti-jamming wireless communications
\cite{WangInfo11}\cite{ICNP11}\cite{QianJSAC12}, short-path routing \cite{Routing07}\cite{Ting13}, non-stochastic channel access affected by primary user
  activity in cognitive radio networks \cite{XYinfocom11} and power control and channel selection problems \cite{MaghTVT2014}.

 The  stochastic and adversarial MABs have co-existed in parallel for almost two decades. Only until recently, the attempt of \cite{BubeckO12} to bring them together did not make it in a full
 sense of unification, since the algorithm relies on the knowledge of time horizon and makes an one-time irreversible switch between
 stochastic and adversarial operation modes if the beginning of the play exhibits adversarial behavior. The first  practical
 algorithm for both stochastic and adversarial bandits is proposed in \cite{Seldin14}. Our current work uses the idea of introducing the novel exploration parameter $\xi_t(f)$ \cite{Seldin14} into our own special combinatorial exponentially weight algorithm by exploiting the channel
 dependency among different strategies.

 This new framework avoids the
   computational inefficiency issue for general combinatorial adversary bandit problems as indicated in  \cite{XYICDCS14} \cite{Lugosi12}. It achieves
  a regret bound of order $O({k_r}\sqrt {tn\ln n})$, which only has a factor of $O(\sqrt{k_r})$ factor off when compared to the optimal
  $O(\sqrt {{k_r}tn\ln n})$ bound in the combinatorial adversary bandit setting \cite{OR2014}. However, we do believe that the regret bound
  in our framework is the optimal one for the exponential weight (e.g. EXP3 \cite{non_MAB02}) type of algorithm settings in the sense that
   the algorithm is computationally  efficient. Thus, our work is also a first computationally efficient combinatorial MAB
    algorithm for general unknown environment\footnote{ As noticed,
    the stochastic combinatorial bandit problem does not have this issue as indicated in \cite{XYICDCS14}\cite{Branislav2015}.}.
 What is more surprising and encouraging, in the stochastic regimes (including the contaminated stochastic regime), all of our algorithms achieve a regret
 bound of order $\tilde{O}(\frac{{n{k_r}\log {{(t)}}}}{\Delta })$. In the sense of
  channel numbers $n$ and size of channels within each strategy $k_r$, this is the best result  for combinatorial
  stochastic bandit problems \cite{Branislav2015}. Please note that
  in \cite{Yi11}, they have a regret bound of order ${O}(\frac{{n^4\log {{(t)}}}}{\Delta })$; in \cite{Liu12}, the
  regret bound is ${O}(\frac{{n^3\log {{(t)}}}}{\Delta })$; and in \cite{Dani08}, the regret bound is ${O}(\frac{{n^2\log^3 {{(t)}}}}{\Delta })$. Thus, our
  proposed algorithms are order optimal with respect to $n$ and ${k_r}$ for all different regimes, which indicate the good
  scalability for general wireless communication systems.

\section{Problem Formulation}
\subsection{System Model}
We consider two wireless devices communicating in an unknown environment, each is
within the other device's transmission range. The sender transmits data packets to the
receiver synchronically  over time. The wireless environment is
 highly flexible in dynamics, where states of the channels could follow different stochastic distributions and could also suffer from
 different kinds of potentially  adversarial attacks. They can also vary over different time slots and channel sets. Without loss of generality (w.l.o.g.), we consider the jamming attack as the representative adversarial model.
 Specifically, we will
 categorize the feature of wireless communication environments
 into four typical regimes in our next discussions. The transceiver pair selects multiple channels to  send and receive signals over a
 set of $n$ available orthogonal channels with
  possibly different data rates across them. We do not differentiate
  channels and frequencies in our discussion.
  During each time slot, the transmitter chooses $k_t$ out of $n$ channels to  transmit data  and the receiver chooses $k_r$  out of $n$ channels to receive data. We assume the transmitter and receiver
  do not pre-share any secrets with each other before data communication, and there is no feedback channel from the receiver to the transmitter. We assume  one jammer launches attack to the transceiver pair over $n$ channels, and the jammer does not
  have the knowledge about the transceiver's strategies before data communication. The data packets rate at time slot $t$ from the transmitter
   on
  channel $f$ is denoted by $g_t(f)$, $g_t(f) \in [0,M]$. Here constant $M$ is the maximum data rate for all channels.


%

\subsection{The Adaptive Uncoordinated Frequency Hopping Problem}
Since no  secret is shared and no adversarial event is informed to the transceiver pair,  the multi-channel wireless communications in unknown environments are  necessary to use frequency hopping strategies to dynamically select a subset of channels to maximize its accumulated data rates over time.  We name ours as the {\emph{Adaptive}} Uncoordinated Frequency Hopping (AUFH) protocol due to its flexibility to achieve optimal performance in various scenarios, when compared to the recent and sophisticated developed UFH protocol in \cite{WangInfo11}\cite{QianJSAC12}. Here  the receiver's selection  of the frequency hopping
strategy to maximize the cumulated data packets reception has the following challenge: 1)  it does not know
the transmitter and adversarial events in the environment, thus it has no good channel access strategy to begin with;
2) the receiver is desirable to have an  adaptively optimal channel access strategy in all different situations.

We consider the  AUFH problem as a sequential
decision problem, where the choice of receiving channels at each time slot is a decision. Denote $\{0,1\}^n$
 as the vector space of all $n$ channels. The strategy space for the transmitter
 is denoted as $S_t \subseteq \{0,1\}^n$ of size $\binom{n}{k_t}$, and the receiver's
 is denoted as $S_r \subseteq \{0,1\}^n$ of size $\binom{n}{k_r}$. If the $f^{th}$-channel is selected for
 transmitting and receiving data, the value of the $f$-th entry of a vector (channel access strategy) is
 $1$, and 0 otherwise. In the case of the existence of jamming attack on a subset of $k_j$ channels, the
  strategy space for the jammer is denoted as $S_j \subseteq \{0,1\}^n$ of size $\binom{n}{k_j}$. For convenience,
  we say that the $f$-th channel is \emph{jammed} if the value of $f$-th entry is 0 and otherwise is 1.  At each time slot, after choosing a strategy $s_r$, the value of the data rate (or called ``reward") $g_t(f)$ is revealed
to the receiver if and only if $f$ is chosen as a receiving channel.

\begin{figure*}
\centering
\includegraphics[scale=.6]{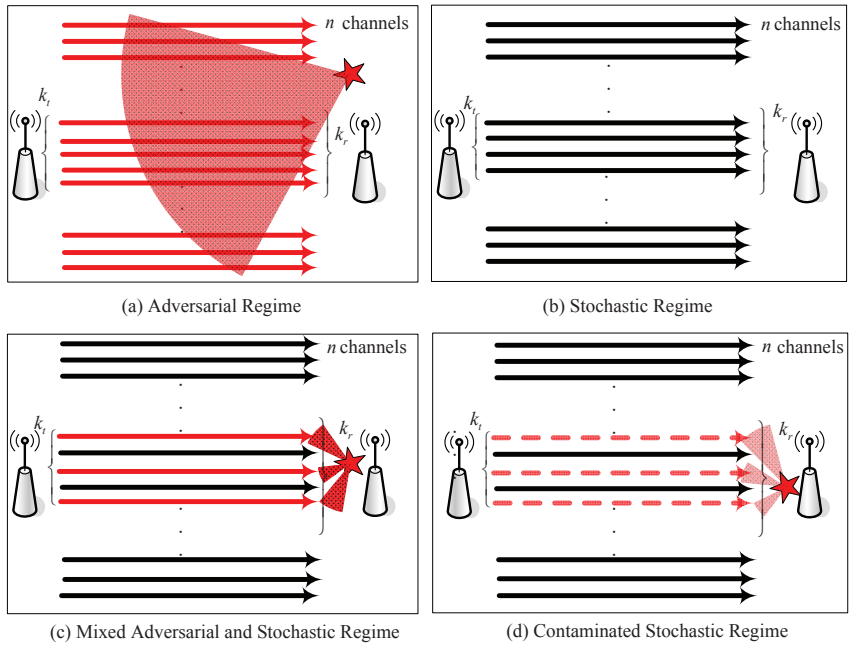}
\caption{Multi-channel Wireless Communications in Different Regimes}
\label{fig:digraph}
\end{figure*}

Formally, the frequency hopping multi-channel access game can be formulated as a MAB problem that is described
as follows: at each time slot $t=1,2,3,...$, the receiver as a decision maker select a strategy $I_t$ from $S_r$. The cardinality of
$S_r$ is $|S_r|=N$. The reward  $g_t(f)$ is assigned to
each channel $f \in \{1,...,n\}$ and the receiver only get rewards in strategy $i \in S_r$. Note that $I_t$ denotes a particular strategy
chosen at time slot $t$ from the receiver's strategy set $S_r$, and $i$ denotes a strategy in $S_r$. The total reward of a strategy
$i$ in time slot $t$ is ${g_t(i)} = \sum\nolimits_{f \in i} {{g_t(f)}}$. Then, on the one hand, the cumulative reward up to time slot $t$ of the strategy
$i$ is $
{G_t(i)} = \sum\nolimits_{s = 1}^t {{g_s(i)}}  = \sum\nolimits_{f \in i} {\sum\nolimits_{s = 1}^t {{g_s(f)}} } .$
On the other hand, the total reward over all the chosen strategies by the receiver up to time slot $t$ is
${\hat G_t} = \sum\nolimits_{s = 1}^t {{g_{s}(I_s)}}  = \sum\nolimits_{s = 1}^t {\sum\nolimits_{f \in {I_s}} {{g_s(f)}} }$,
where the strategy $I_s$ is chosen randomly according to some distribution over $S_r$. The performance of this algorithm
is qualified by \emph{regret} $R(t)$, defined as the difference between the expected number of successfully received data packets
using our proposed algorithm and the expected rewards that use the best fixed solution up to $t$ time slots for the game, i.e.,
 \begin{IEEEeqnarray*}{l}
R(t) = \mathop {\max }\limits_{i \in {S_r}} \mathbb{E} \left\{ {{G_t(i)}} \right\} - \mathbb{E}\left[ {{{\hat G}_t}} \right],
\IEEEyesnumber
\end{IEEEeqnarray*}
where the maximum is taken over all available strategies to the receiver. However, during the theoretical analysis of the AUFH-based algorithm
in our next discussion, if we
 use the \emph{gain} (reward) model, we have to apply additional smoothing of the playing distribution $q_t(f)$ regarding $\tilde g_t(f)$. Thus,  we can introduce the \emph{loss} model by the simple trick of $  \ell_t(f)= 1- g_t(f)$ for each channel $f$ and $\ell_t(i)= k_r -g_t(i)$ for
 each strategy, respectively. Then, we have $ L_t(i) = tk_r - G_t(i)$ where ${L_t}(i) = \sum\nolimits_{s = 1}^t \ell_t(i)=  {\sum\nolimits_{s = 1}^t {
 \sum\nolimits_{f \in i} {{\ell_s(f)}} } }$, and similarly, we have $ \hat{L}_t = tk_r - \hat{G}_t$. We  use $\mathbb{E}_t [\cdot]$ to denote expectations on
 realization of all strategies as random variables up to round $t$.  Therefore, the expected regret $R(t)$ can be rewritten as
  \begin{IEEEeqnarray*}{l}
  \begin{array}{l}
\!\!\!\!\!\!\!\!\!  \mathbb{E}\left[ {{{\hat L}_t}} \right] - \mathop {\min}\limits_{i \in {S_r}} \mathbb{E} \left\{ {{L_t(i)}} \right\}
= \mathbb{E} \sum\limits_{s = 1}^t {{\ell_t}({I_t})}  - \mathop {\min }\limits_{i \in {S_r}} \mathbb{E} \sum\limits_{s = 1}^t {{\ell_t}(i)} \\
   =\mathbb{E} [ \sum\limits_{s = 1}^t {\mathbb{E}_s [ {\sum\limits_{f \in {I_s}} {{\ell_s(f)}} } ]}]  - \mathop {\min }\limits_{i \in {S_r}} ( { \mathbb{E}   [  {\sum\limits_{s = 1}^t
 \mathbb{E}_s [ {\sum\limits_{f \in i} {{\ell_s(f)}}  }] }\!\! ]} ).
 \end{array}\IEEEyesnumber \label{eq:Regrets}
\end{IEEEeqnarray*}
The expectation is taken over the possible randomness of the proposed algorithm and loss generation model. The goal of the algorithm is
to minimize the regret. The above definition of regret is usually named as the \emph{pseudo regret} \cite{Bubeck12}, which is upper
bounded by the expected regret $\mathbb{E} \{ \bar R(t)\} = \mathbb{E} \{ \sum\nolimits_{s = 1}^t {{\ell_t}({I_t})}  -
\mathop {\min }\nolimits_{i \in {S_r}} \sum\nolimits_{s = 1}^t {{\ell_t}(i)} \}$. Only when the adversary is oblivious, who prepares the entire
sequence of loss functions ${\ell_t}({I_t})$ ($t=1,2,3,...$) in advance,  pseudo regret (\ref{eq:Regrets}) coincides with the standard
expected regret $\mathbb{E} \{ \bar R(t)\}$\cite{Bubeck12}.

Note that the choice of the loss function at time slot $t$ of the oblivious adversary is
independent to the first $t-1$ time slots. Otherwise, the adversary can be called an \emph{adaptive adversary}. In this case, let us
denote the decision maker's entire sequence of strategies up to current timslot $t$ as $(I_1,..., I_t)$, which we abbreviate by $I_{1,...,t}$. The
\emph{expected cumulative loss} suffered by the player after $t$ rounds is $\mathbb{E}[
 {\sum\nolimits_{s = 1}^t {{\ell_s}({I_{1,...,s}})} } ]$. We need to compare it with a \emph{competitor
class} $\mathcal{C}_t$, which is simply a set of deterministic strategy sequences of length $t$. Intuitively, we
 would like to compare the decision maker's loss with the cumulative loss of the best action sequence in $\mathcal{C}_t$. In practice, the
most common way to evaluate the decision maker's performance is to measure its \emph{external pseudo-regret} compared to $\mathcal{C}_t$ \cite{non_MAB02}.
Thus, the
regret for adaptive adversary is defined as,
  \begin{IEEEeqnarray*}{l}
 \! \! \! \!\!R(t) \!\!= \mathop {\max }\limits_{({y_1},...,{y_t})
\in {\mathcal{C}_t}} \mathbb{E} [ \! {\sum\limits_{s = 1}^t {( {{\ell_s}({I_{1,...,s}}) - {\ell_s}({I_{1,...,s - 1}},{y_s})} )} } ].
\IEEEyesnumber \label{eq:Regrets2}
\end{IEEEeqnarray*}
This regret definition is suitable for most of the theoretical works of the online learning and bandit setting. If the
adversary is oblivious, we have ${{\ell_t}({I_{1,...,t}})}$ equals to ${{\ell_t}({I_{t}})}$. With this simplified notation, the
regret in (\ref{eq:Regrets2}) becomes
  \begin{IEEEeqnarray*}{l}
\mathds{E}[ {\sum\limits_{s = 1}^t {{\ell_s}({I_s})} }
 ] - \mathop {\min }\limits_{({y_1},...,{y_t}) \in {\mathcal{C}_t}} \sum\limits_{s = 1}^t {{\ell_s}({y_s})},
\end{IEEEeqnarray*}
which is exactly the same as (\ref{eq:Regrets}) with $\mathcal{C}_t= S_r$, if we take an expectation over all the strategy
sequence $({y_1},...,{y_t})$.

\subsection{The Four Regimes of Wireless Environments}
Since our algorithm does not need to know the nature of the environments, there exist  different features of  the environments
that will affect its performance. We categorize them into the four typical regimes as shown in Fig. 1.

\subsubsection{Adversarial Regime}
In this regime, there is a jammer sending interfering power or injecting garbage data packets
over all $n$ channels such that the transceiver's channel rewards are completely suffered by an unrestricted jammer (See Fig.1 (a)). When we assume the use of the same level of
transmission power as in the stochastic regime, the data rate will be significantly reduced in the adversarial regime. Note that, as a classic  model of the well known non-stochastic MAB problem \cite{non_MAB02}, the adversarial regime implies that the jammer often launches attack in almost\footnote{
 Strictly speaking, according to the definition and analysis of the contaminated stochastic regime in the next discussion, when the total number of
 contaminated locations of round-channel pairs $(t,f)$ by the jammer on each
 channel up to time $t$ is largely great than $t\Delta(f)/4$ on average, then we can regard it belongs to the adversarial regime.} every time slot. It is the most general setting and the other three regimes can be regarded as  special cases of the
  adversarial regime. Obviously, a  strategy $i \in \mathop {\arg \min }\nolimits_{i' \in {S_r}} \{ {\mathbb{E}[ {\sum\nolimits_{s = 1}^t {{L_t}(i')} } ]} \}$ is known as a \emph{ best strategy in hindsight} for the first $t$ round.

\textbf{Attack Model:} Different attack philosophies will lead to different level of effectiveness. We classify  various types of jammers into the following
two categories in the adversarial regime:¡¡

  a) \emph{Oblivious jammer:} an oblivious jammer could attack different channels with different jamming strength as a result of different
data rate reductions. Its current attacking
strategy is not based the observed past communication records. As described in \cite{QianJSAC12}, oblivious jammer
can use \emph{static} and \emph{random} strategies to attack wireless channels. If its attacking strategy is time-independent (e.g. static jammer),
we can simply regard it as a stochastic channel with bad channel quality. Usually, the attacking strategy for oblivious jammer can change with time. As noticed, many other
 kinds of jammers, such as partial band jamming, sweep jamming etc. \cite{Jamming09} all belong to the oblivious attack model. Briefly, it is  a simple
attack model that does not react to the defending algorithm, although the attackers' attacking strategies could be largely different.



 b) \emph{Adaptive jammer:} an adaptive jammer, also named as non-oblivious jammer,  adaptively selects its jamming strength on the targeted (sub)set of jamming channels by utilizing its
 past experience and observation of the previous communication records.
  In the adversarial regime, we consider that the adaptive jammer is very powerful in the sense that it does not only know the
  communication protocol and able to attack with different level of strength over a subset of channels for data communications during a
  single time slot, but it also can monitor all the $n$ available channels during the same time slot. For example, the reactive
   jammer with the behavior described in \cite{Jamming09} belongs to this type. As shown in a recent work\cite{Arora12}, no bandit algorithm can guarantee a sublinear
regret $o(t)$ against an adaptive adversary with unbounded memory. The adaptive adversary can mimic and perform the same
learning algorithm as the decision maker, i.e., the receiver in our work. It can set the same channel access probabilities
as the channel access algorithm, which will lead to a linear regret. Therefore, we consider a more practical \emph{m-memory-bounded adaptive adversary} \cite{Arora12} model, which is constrained
to choose loss functions that depend only on the $m+1$ most recent strategies.

\subsubsection{Stochastic Regime}
In this regime, the transceiver communicates over $n$ stochastic channels as shown in Fig.1 (b). The channel losses $\ell_t(f), \forall f \in 1,...,n$ (Obtained
by transferring  the reward to loss $\ell_t(f)=1- g_t(f)$) of each channel $f$ are sampled independently from an unknown distribution that depends on $f$, but not on $t$. We use $\mu_f = \mathbb{E}
\left[ {{\ell_t(f)}} \right]$ to denote the expected loss of channel $f$. We define channel $f$ as the \emph{best channel}
if $\mu (f) = {\min _{f'}}\{ {\mu (f')} \}$ and \emph{suboptimal channel} otherwise; let $f^*$ denote some best channel. Similarly,  for each strategy
$i \in S_r$, we have  the \emph{best strategy} $\mu (i) = {\min _{i'}}\{ {\sum\nolimits_{f \in i'} {\mu (f)} } \}$ and \emph{suboptimal
strategy} otherwise; let $i^*$ denote some best strategy. For each channel $f$, we define the gap $\Delta (f) = \mu (f) - \mu ({f^*})$;
let $\Delta_f = {\min _{f:\Delta (f) > 0}}\left\{ {\Delta (f)} \right\}$ denote the minimal gap of channels, or the
gap from the \emph{second} best channel(s). Similarly, for each strategy $i$, we have
 $\Delta (i) = \mu (i) - \mu ({i^*})$; let $\Delta_i  = {\min _{i:\Delta (i) > 0}}\left\{ {\Delta (i)} \right\}$ denote the minimal gap of strategies. Let $N_t(f)$ and $N_t(i)$ be the respective number of times channel $f$ and strategy $i$ was played up to time $t$, the regret can be rewritten as $\tilde R(t)$ based on  $N_t(f)$, and we have
  \begin{IEEEeqnarray*}{l}
R(t) = \sum\limits_i {\mathbb{E}\left[ {{N_t}(i)} \right]} \Delta (i) \le \tilde R(t) = \sum\limits_f {\mathbb{E}\left[ {{N_t}(f)} \right]} \Delta (f).
\IEEEyesnumber \label{eq:StocR}
\end{IEEEeqnarray*}
Note that we calculate the upper bound regret $\tilde R(n)$ from the perspective of channel set $\mathcal{K}$, where the regret is upper bounded by
the regret from the perspective of strategies set $\mathcal{N}$.  This is because the set of strategies is of the size  $\binom{n}{k_r}$ that grows exponentially with respect to $n$ and it does not exploit the channel dependency among different strategies. We thus calculate the upper regret from the perspective of channels, where tight regret bounds are achievable.

\subsubsection{Mixed Adversarial and Stochastic Regime}
This regime assumes that the jammer only attack $k_j$ out $n$ channels at each time slot.
As shown in Fig.1 (c),  there is always a
$k_j/n$ portion of channels that  suffers from jamming attack while the other $(n-k_j)/n$ portion is stochastically distributed. We
call this regime  the mixed adversarial and stochastic regime.

\textbf{Attack Model:} We consider the same type of jammer as described in the adversarial regime for the mixed adversarial and stochastic regime, which
includes: static jamming and random jamming of the oblivious jammer and the adaptive jammer. The difference here is that the jammer
only attacks a subset of channels of size $k_j$ over the total  $n$ channels not all channels.


\subsubsection{Contaminated Stochastic Regime}
The definition of the contaminated stochastic regime comes from many practical observations that only a few channels and time slots are exposed to adversary. Here comes the question: is this
environment still stochastic or adversarial?  We are fortunate to answer this question. In this regime, for oblivious jammer, it selects some slot-channel pairs $(t,f)$ as ``locations" to attack before the multi-channel wireless communications start, while the remaining channel rewards are generated the same as the stochastic regime. We can introduce and define
the \emph{attacking strength} parameter $\zeta \in [0, 1/4)$. After certain $\tau$ timslots, for all $t> \tau$ the total number of contaminated
locations of each suboptimal channel up to time $t$ is $t\Delta(f)\zeta$ and the number of contaminated locations of each best
channel is $t\Delta_f \zeta$.

We call a contaminated stochastic regime \emph{moderately contaminated}, if by the definition $\zeta$ is at most $1/4$, we can prove that for all $t> \tau$ on the average over the stochasticity of the
loss sequence the adversary can reduce the gap of every channel by at most one half. Thus, if the attacking strength $\zeta \in [0, 1/4]$, the
environment can still be regarded as benign that behaves stochastically (though it is contaminated).

\begin{algorithm}
\caption{AUFH-EXP3++: An MAB-based Algorithm for AUFH}
\begin{algorithmic}
\STATE \textbf{Input}: $n, k_r, t$,  and See text for definition of $\eta_t$ and $\xi_t(f)$.
\STATE \textbf{Initialization}: Set initial channel and strategy losses $\forall i \in [1, N], \tilde{L}_0(i)= 0$ and $\forall f \in [1, n], \tilde{\ell}_0(f)= 0$, respectively; Then the initial channel and strategy weights $\forall i \in [1, N],
W_0(i)= k_r$ and $\forall f \in [1, n], w_0(f)= 1$, respectively. The initial total strategy weight $W_0=N=\binom{n}{k_r}$.
\!\STATE \textbf{Set}: \!\! $\beta_t \!\! =\!\! \frac{1}{2}\sqrt {\frac{{\ln n}}{{tn}}}$; ${\varepsilon _t}\left( f \right) \!= \! \min \left\{ {\frac{1}{{2n}},{\beta _t},{\xi _t}\left( f \right)} \right\},\forall f \in \left[ {1,n} \right]$.
\FOR { time slot $t=1,2,...$}
\STATE 1: The receiver selects a channel hopping strategy $I_t$ at random according to the strategy's probability $p_t(i),\forall f \in \left[ {1,n} \right]$, with $p_t(i)$ computed as follows:
\begin{IEEEeqnarray*}{l}{p_t}(i) = \left\{ \begin{array}{l}
\!\!\!  \left( {1 - \sum\nolimits_{f=1}^{n} {{\varepsilon _t}(f)} } \right)\frac{{{w_{t - 1}}\left( i \right)}}{{{W_{t - 1}}}} + \! \sum\limits_{f \in i} {{\varepsilon _t}(f)} \ \emph{if} \ i \in \mathcal{C} \\
\!\!\!  \left( {1 - \sum\nolimits_{f=1}^{n}  {{\varepsilon _t}(f)} } \right)\frac{{{w_{t - 1}}\left( i \right)}}{{{W_{t - 1}}}} \  \quad   \quad \quad \quad \
\text{\emph{if}} \ i \notin \mathcal{C}
\end{array} \right.
\end{IEEEeqnarray*}
The computation is taken for the probability distributions over all strategies $p_t=(p_t(1),p_t(2),...,p_t(N))$.

\STATE 2: The receiver computes the probability $q_t(f),\forall f \in \left[ {1,n} \right]$, as
\begin{IEEEeqnarray*}{l}
\begin{array}{l}
{q_t}(f) = \sum\limits_{i:f \in i} {{p_t}(i)}
 = \left( {1 - \sum\nolimits_{f=1}^{n} {{\varepsilon _t}(f)} } \right)\frac{{\sum\nolimits_{i:f \in i} {{w_{t - 1}}\left( i \right)} }}{{{W_{t - 1}}}} \\
\quad \quad\quad \quad \quad \quad \quad \quad \ \ + \sum\limits_{f \in i} {{\varepsilon _t}(f)}
 \left| {\left\{ {i \in \mathcal{C}:f \in i} \right\}} \right|.
\end{array}
\end{IEEEeqnarray*}
Then, the probability distributions over all channels are $q_t=(q_t(1),q_t(2),...,q_t(n))$.

\STATE 3: The receiver calculates the loss for channel $f$, $\ell_{t-1}(f), \forall f \in I_t$ based on the received
channel gain $g_{t-1}(f)$ by using $\ell_{t-1}(f) = 1- g_{t-1}(f)$. Compute the estimated loss $\tilde{\ell}_{t}(f), \forall f \in [1,n]$ as follows:
\begin{IEEEeqnarray*}{l}
{\tilde{\ell}_t}(f) = \left\{ \begin{array}{l}
\frac{{{\ell_t}(f)}}{{{q_t}(f)}} \quad  \text{\emph{if channel}}  f \in {I_t} \\
0         \ \ \quad\quad \emph{otherwise}.
\end{array} \right.
\end{IEEEeqnarray*}

\STATE 4: The receiver updates all the weights as
\begin{IEEEeqnarray*}{l}
\begin{array}{l}
\begin{array}{l}
{w_t}\left( f \right) = {w_{t - 1}}\left( f \right){e^{ - \eta_t { {\tilde \ell}_t}(f)}} = {e^{ - \eta_t {{\tilde L}_t}(f)}}.\\
{{\bar w}_t}\left( i \right) = \prod\limits_{f \in i} {{w_t}(f)}  = {{\bar w}_{t - 1}}\left( i \right){e^{ - \eta_t {{\tilde \ell}_t}(i)}}.
\end{array}
\end{array}
\end{IEEEeqnarray*}
where ${{\tilde L}_t}(f) = {{\tilde L}_{t - 1}}(f) + {{\tilde \ell}_{t - 1}}(f),{{\tilde \ell}_{t - 1}}(e) = \sum\nolimits_{f \in i} {{{\tilde \ell}_{t - 1}}(f)}$ and ${{\tilde L}_t}(i) = {{\tilde L}_{t - 1}}(i) + {{\tilde \ell}_{t - 1}}(i)$. The sum of the total weights of the strategies is
\begin{IEEEeqnarray*}{l}
{W_t} = \sum\limits_{i \in S_r} {{{\bar w}_t}\left( i \right)}
\end{IEEEeqnarray*}
\ENDFOR
\end{algorithmic}
\end{algorithm}
\vspace{-.2cm}

\section{The Optimal Adaptive Uncoordinated Frequency Hopping Algorithm}
In this section, we develop an AUFH algorithm in the receiver side. The design philosophy is that the receiver  collects and learns the rewards of the
previously chosen channels, based on which it can decide the next time slot channel access strategy.   The main difficulty is that the algorithm is required
 to appropriately balance between \emph{exploitation} and
\emph{exploration}. On the one hand, the algorithm needs to keep exploring the best set of channels to receive the data packets due to the dynamic
changing of the environments; on the other hand, the algorithm needs to exploit the already selected best set of channels so that
they will not be under-utilized. 

We describe the Algorithm 1 named as AUFH-EXP3++. It  is a variant based on
EXP3 algorithm and \cite{QianJSAC12}, whose performance in the four regimes will be asymptotically
optimal. Our new algorithm uses the fact that when rewards of channels of the chosen strategy are revealed as in step $1$ of the Algorithm 1, this
also provides some information about  the rewards of each strategy sharing common channels with the chosen strategy, i.e.,  the probability that all the strategies that share the same channel would be projected to it in step $2$.  As noticed,  the
conversion from rewards (gains) to losses is done  to facilitate subsequent performance analysis. During each time slot, we assign
a channel weight that is dynamically adjusted based on the channel losses revealed to the receiver as shown in step $3$. Then, in the step $4$, the weight of a strategy is determined by the product of weights of all channels.

 Compared to \cite{QianJSAC12} that targets only for \emph{secure} wireless communications, our algorithm has two control parameters: the \emph{learning rate} $\eta_t$ and the exploration
parameters $\xi_t(f)$ for each channel $f$, whereas the algorithm in \cite{QianJSAC12} does not explore the using of the parameters $\xi_t(f)$ to
detect the other regimes of the environment. The key innovation here is that we have used the advanced
\emph{martingale concentration inequalities} (i.e., Lemma 8) to detect i.i.d, contaminated and non-i.i.d. behaviors without the knowledge about the nature of the environments, and the exploration parameter  $\xi_t(f)$  is tuned individually for each channel depending on the past observations.

Let $N$ denote the total number of strategies at the receiver side. A set of \emph{covering strategy} is defined to ensure that
each channel is sampled sufficiently often. It has the property,  for each channel $f$, there is a strategy $i \in \mathcal{C}$ such that
$f \in i$. Since there are only $n$ channels and each strategy includes $k_r$ channels, we have $|\mathcal{C}| = \lceil {\frac{n}{{{k_r}}}} \rceil$.
 The value $ \sum\nolimits_{f \in i} {{\varepsilon _t}(f)}$ means the randomized
 exploration probability for each strategy $i \in \mathcal{C}$, which is the summation of each channel $f$'s exploration probability ${\varepsilon _t}\left( f \right)$ that belongs to the strategy $i$. The introduction of $\sum\nolimits_{f \in i} {{\varepsilon _t}\left( f \right)}$ ensures that $p_t(i) \ge  \sum\nolimits_{f \in i} {{\varepsilon _t}(f)}$ so that it is a mixture of exponentially
 weighted average distribution and uniform distribution  \cite{Auer95} over each strategy.

In the following discussion,  the learning rate $\eta_t$ is sufficient to control and obtain the
regret of the AUFH-EXP3++ in the adversarial regime, regardless of the choice of exploration parameter $\xi_t(f)$. The exploration parameter $\xi_t(f)$ is sufficient to control the regret of AUFH-EXP3++ in the stochastic regimes regardless of the choice of
 $\eta_t$, as long as $\eta_t \ge \beta_t$. To facilitate the AUFH-EXP3++ algorithm without knowing about the nature of environments, we can apply the
 two control levers simultaneously by setting $\eta_t = \beta_t$ and use the control parameter $\xi_t(f)$ in the stochastic regimes such that it can achieve
 the optimal ``root-t" regret in the adversarial regime and almost optimal ``logarithmic-t" regret in the stochastic regime (though with a suboptimal
 power in the logarithm).


\section{Performance Results in Different Regimes}
We analyze the regret performance of our proposed AUFH-EXP3++ algorithm in different regimes in the following section. W.l.o.g., we normalize
$M=1$ in all our results to facilitate clear comparisons with regret bounds of others' works.

\subsection{Adversarial Regime}
We first show that tuning $\eta_t$ is sufficient to control the regret of AUFH-EXP3++ in the adversarial regime, which is a
general result that holds for all other regimes.

\textbf{Theorem 1.}  Under the \emph{oblivious} jamming attack, no matter how the status of the channels change (potentially in an adversarial manner), for $\eta_t = \beta_t$ and any
$\xi_t(f) \ge 0$, the regret of the AUFH-EXP3++ algorithm for any $t$ satisfies:
\begin{IEEEeqnarray*}{l}
R(t) \le 4{k_r}\sqrt {tn\ln n}.
\end{IEEEeqnarray*}

\textbf{Theorem 2.}  Under the \emph{m-memory-bounded adaptive} jamming attack, no matter how the status of the channels change (potentially in an adversarial manner), for $\eta_t = \beta_t$ and any
$\xi_t(f) \ge 0$, the regret of the AUFH-EXP3++ algorithm for any $t$ is upper bounded by:
\begin{IEEEeqnarray*}{l}
R(t) \le (m + 1){(4{k_r}\sqrt {n\ln n} )^{\frac{2}{3}}}{t^{\frac{2}{3}}} + o({t^{\frac{2}{3}}}).
\end{IEEEeqnarray*}

\subsection{Stochastic Regime}
Now we show that for any $\eta_t \ge \beta_t$, tuning the exploration parameters $\xi_t(f)$ is sufficient to control the regret of the algorithm
in the stochastic regime. We consider a different number of ways of tuning the exploration parameters $\xi_t(f)$ for different practical
 implementation considerations, which will lead to different regret performance of AUFH-EXP3++. We begin with an idealistic assumption that the gaps $\Delta(f), \forall f \in n$ is known, just to
give an idea of what is the best result we can have and our general idea for all our proofs.


\textbf{Theorem 3}. Assume that the gaps $\Delta(f), \forall f \in n,$  are known. Let $t^*$ be the
minimal integer that satisfy ${t^*(f)} \ge \frac{{4{c^2}n\ln {{({t^*(f)}\Delta {{(f)}^2})}^2}}}{{\Delta {{(f)}^4}\ln (n)}}$.
For any choice of $\eta_t \ge {\beta _t}$
 and any $c \ge 18$, the regret of the AUFH-EXP3++ algorithm with $\xi_t(a)=\frac{{c\ln (t\Delta {{(f)}^2})}}{{t\Delta {{(f)}^2}}}$ in the
 stochastic regime satisfies:
   \begin{displaymath}
   \begin{array}{l}
 R(t) \le \sum\limits_{f= 1, \Delta {(f)} >0  }^n {O\left( {\frac{{k_r\ln {{(t)}^2}}}
 {{\Delta {{(f)}}}}} \right)}  + \sum\limits_{f= 1, \Delta {(f)} >0  }^n \Delta {{(f)}} t^* {{(f)}}\\
\hspace{.7cm} = {O\left( {\frac{{k_r n\ln {{(t)}^2}}} {{\Delta_{f} {{}}}}} \right)}
+ \sum\limits_{f= 1, \Delta {(f)} >0  }^n \tilde O\left( {\frac{n}{{\Delta {{(f)}^3}}}} \right).
 \end{array}
   \end{displaymath}
From the upper bound results, we note that the leading constants $k_r$  and $n$ are optimal and tight as indicated in CombUCB1 \cite{Branislav2015} algorithm. However, we have a factor of $\ln(t)$ worse of the regret performance than
the optimal ``logarithmic" regret as in \cite{Robbins1985}\cite{Branislav2015}.


\subsubsection{A Practical Implementation by estimating the gap}
Because of the gaps $\Delta(f), \forall f \in n$ can not be known in advance before running the algorithm.
In the next, we show a more practical result that using the empirical gap as an estimate
of the true gap. The estimation process can be performed in background for each channel $f$ that
starts from the running of the algorithm, i.e.,
 \begin{displaymath}
{{\hat \Delta }_t}(f) = \min \left\{ {1,\frac{1}{t}\left({{\tilde L}_t}(f) -\mathop {\min }\limits_{f'} ({{\tilde L}_t}(f'))\right )} \right\}.
\IEEEyesnumber \label{eq:EstD}
 \end{displaymath}
 This is a first algorithm that can be used in many real-world applications.

\textbf{Theorem 4.} Let $c \ge 18$ and $\eta_t \ge \beta_t$. Let $t^*$ be the minimal integer that satisfies $t^*
 \ge \frac{{4{c^2}\ln {{(t^*)}^4} n}}{{\ln (n)}}$, and let ${t^*}(f) = \max \left\{ {{t^*},\left\lceil {{e^{1/\Delta {{(f)}^2}}}} \right\rceil } \right\}$ and
$t^*=max_{\{f \in n\}} t^* {{(f)}}$. The regret of the AUFH-EXP3++ algorithm
with ${\xi _t}(f) = \frac{{c{{\left( {\ln t} \right)}^2}}}{{t{{\hat \Delta }_{t-1}}{{(f)}^2}}}$, termed as
AUFH-EXP3++$^\emph{AVG}$, in the stochastic regime satisfies:
   \begin{displaymath}
      \begin{array}{l}
 R(t) \le \sum\limits_{f= 1, \Delta {(f)} >0  }^n {O\left( {\frac{{k_r\ln {{(t)}^3}}}
 {{\Delta {{(f)}}}}} \right)}  + \sum\limits_{f= 1, \Delta {(f)} >0  }^n \Delta {{(f)}} t^* {{(f)}}\\
 \hspace{.7cm}=O\left( {\frac{{n k_r\ln {{(t)}^3}}}
 {{\Delta_{f} }}} \right) + n t^*.
    \end{array}
   \end{displaymath}
From the theorem, we see in this more practical case,  another factor of $ln(t)$ worse of the regret performance when compared to the idealistic
case. Also, the additive constants $t^*$ in this theorem can be  very large. However, our experimental results show that
a minor modification of this algorithm performs comparably to ComUCB1 \cite{Branislav2015} in the stochastic regime.


%


\subsection{Mixed Adversarial and  Stochastic Regime}
The mixed adversarial and stochastic regime can be regarded as a special case of mixing
adversarial and stochastic regimes. Since there is  always a jammer randomly attacking $k_j$ channels
constantly over time,  we will have the following theorem for the AUFH-EXP3++$^\emph{AVG}$ algorithm, which is
a much more refined regret performance bound than the general regret bound in the adversarial regime.

\textbf{Theorem 5.} Let $c \ge 18$ and $\eta_t \ge \beta_t$. Let $t^*$ be the minimal integer that satisfies $t^*
 \ge \frac{{4{c^2}\ln {{(t^*)}^4} n}}{{\ln (n)}}$, and Let ${t^*}(f) = \max \left\{ {{t^*},\left\lceil {{e^{1/\Delta {{(f)}^2}}}} \right\rceil } \right\}$ and
$t^*=max_{\{f \in n\}} t^* {{(f)}}$. The regret of the AUFH-EXP3++ algorithm
with ${\xi _t}(f) = \frac{{c{{\left( {\ln t} \right)}^2}}}{{t{{\hat \Delta }_{t-1}}{{(f)}^2}}}$, termed as
AUFH-EXP3++$^\emph{AVG}$ under \emph{oblivious jamming} attack, in the mixed stochastic and adversarial regime satisfies:
   \begin{displaymath}
      \begin{array}{l}
 R(t) \le \sum\limits_{f= 1, \Delta {(f)} >0  }^{n-k_r} {O\left( {\frac{{k_r\ln {{(t)}^3}}}
 {{\Delta {{(f)}}}}} \right)}  + \sum\limits_{f= 1, \Delta {(f)} >0  }^{n-k_r} \Delta {{(f)}} t^* {{(f)}} \\
 \hspace{.9cm} + 4{k_j}\sqrt {tn\ln n}\\
 \hspace{.7cm}=O\left( {\frac{{{(n-k_j)} k_r\ln {{(t)}^3}}}
 {{\Delta_{f} }}} \right) + n t^* + O\left({k_j}\sqrt {tn\ln n}\right).
    \end{array}
   \end{displaymath}
Note that the results in Theorem 5 has better regret performance than the results obtained by adversarial MAB as shown
 in Theorem 1 and the anti-jamming algorithm in \cite{QianJSAC12}.

\textbf{Theorem 6.} Let $c \ge 18$ and $\eta_t \ge \beta_t$. Let $t^*$ be the minimal integer that satisfies $t^*
 \ge \frac{{4{c^2}\ln {{(t^*)}^4} n}}{{\ln (n)}}$, and Let ${t^*}(f) = \max \left\{ {{t^*},\left\lceil {{e^{1/\Delta {{(f)}^2}}}} \right\rceil } \right\}$ and
$t^*=max_{\{f \in n\}} t^* {{(f)}}$. The regret of the AUFH-EXP3++ algorithm
with ${\xi _t}(f) = \frac{{c{{\left( {\ln t} \right)}^2}}}{{t{{\hat \Delta }_{t-1}}{{(f)}^2}}}$, termed as
AUFH-EXP3++$^\emph{AVG}$ \emph{m-memory-bounded adaptive} jamming attack, in the mixed stochastic and adversarial regime satisfies:
   \begin{displaymath}
      \begin{array}{l}
 R(t) \le \sum\limits_{f= 1, \Delta {(f)} >0  }^{n-k_r} {O\left( {\frac{{k_r\ln {{(t)}^3}}}
 {{\Delta {{(f)}}}}} \right)}  + \sum\limits_{f= 1, \Delta {(f)} >0  }^{n-k_r} \Delta {{(f)}} t^* {{(f)}} \\
   \quad\quad\quad    + (m + 1){(4{k_j}\sqrt {n\ln n} )^{\frac{2}{3}}}{t^{\frac{2}{3}}} + o({t^{\frac{2}{3}}})\\
 \hspace{.7cm}=O\left( {\frac{{{(n-k_j)} k_r\ln {{(t)}^3}}}
 {{\Delta_{f} }}} \right) + n t^* + O\left({({k_j}\sqrt {n\ln n} )^{\frac{2}{3}}}{t^{\frac{2}{3}}}\right).
    \end{array}
   \end{displaymath}
The results shown in Theorem 6 provides the first quantitative regret performance under
adaptive jamming attack, while the related work \cite{QianJSAC12} with the similar adversary model and the
same communication scenario in this case only provided
simulation results demonstrations.

\subsection{Contaminated stochastic regime}
We  show that the algorithm AUFH-EXP3++$^\emph{AVG}$ can still retain ``polylogarithmic-t" regret in the  contaminated stochastic
regime with a potentially large leading constant in the performance. The following is the result for the \emph{moderately contaminated stochastic regime}.


\textbf{Theorem 7.} Under the setting of all parameters given in Theorem 3, for ${t^*}(f) = \max \left\{ {{t^*},\left\lceil {{e^{4
/\Delta {{(f)}^2}}}} \right\rceil } \right\}$, where $t^*$ is defined as before  and
$t_3^*=max_{\{f \in n\}} t^* {{(f)}}$, and the attacking strength parameter $\zeta \in [0,1/2)$ the regret of the \rm{AUFH-EXP3++} algorithm in the
contaminated stochastic regime that is contaminated after $\tau$ steps satisfies:
   \begin{displaymath}
         \begin{array}{l}
\!\!  R(t) \!\! \le \!\! \sum\limits_{f= 1, \Delta {(f)} >0  }^n \!\!\!\!   {O\left( {\frac{{k_r\ln {{(t)}^3}}}
 {{(1-2\zeta)\Delta {{(f)}}}}} \right)} \!\!  + \!\! \!\! \sum\limits_{f= 1, \Delta {(f)} >0  }^n \!\! \!\! \!\!  \Delta {{(f)}} \max\{t^* {{(f)}}, \tau\}. \\
  \hspace{.6cm}= {O\left( {\frac{{n k_r\ln {{(t)}^3}}}
 {{(1-2\zeta)\Delta_{f}}}} \right)} + K t_3^*.
       \end{array}
   \end{displaymath}


If $\zeta \in (1/4, 1/2
)$, we can find that the leading factor $1/(1-2\zeta)$ is very large, which is \emph{severely} contaminated. Now, the obtained regret bound is not quite meaningful, which could be much worse than the regret performance in the adversarial regime for
both oblivious and adaptive adversary.



 \section{Proofs of Regrets in Different Regimes}
 We prove the theorems of the performance results from the previous section in the order they were presented.
  \subsection{The Adversarial Regimes}
 The proof of Theorem 1 borrows some of   the analysis of EXP3 of the loss model in  \cite{Bubeck12}. However, the introduction
 of the new mixing   exploration parameter and the truth of channel/frequency dependency as a special type of combinatorial MAB
 problem in the loss model  makes the proof a non-trivial task, and we prove it for the first time.

 \emph{Proof of Theorem 1.}
  \begin{IEEEproof}
Note first that the following equalities can be easily verified:
${\mathbb{E}_{i \sim {p_t}}}{\tilde{\ell}_t}(i) = {\ell_t}({I_t}),{\mathbb{E}_{{\tilde{\ell}_t} \sim {p_t}}}{\ell_t}(i) = {\ell_t}(i),{\mathbb{E}_{i \sim {p_t}}}{\tilde{\ell}_t}{(i)^2} = \frac{{{\ell_t}{{({I_t})}^2}}}{{{p_t}({I_t})}}$ and ${\mathbb{E}_{{I_t} \sim {p_t}}}\frac{1}{{{p_t}({I_t})}} = N$.


Then, we can immediately rewrite $R(t)$ and have
\begin{IEEEeqnarray*}{l}
R(t)  =\mathbb{E}_t \left[\sum\limits_{s = 1}^t {{\mathbb{E}_{i \sim {p_s}}}{\tilde{\ell}_s}(i)}  - \sum\limits_{s = 1}^t {{\mathbb{E}_{{I_s} \sim {p_s}}}{\tilde{\ell}_s}(i)} \right].
\end{IEEEeqnarray*}

The key step here is to consider the expectation of the cumulative losses ${\tilde{\ell}_t}(i)$ in the sense of distribution
$i \sim {p_t}$. Let ${\varepsilon_t}(i)=\sum\nolimits_{f \in i} {{\varepsilon _t}(f)}$.  However, because of the mixing terms of $p_t$, we need to introduce a few more notations. Let $u =
( {\underbrace {\sum\nolimits_{f \in 1}
{{\varepsilon _t}(f)},...,\sum\nolimits_{f \in i}
{{\varepsilon _t}(f)},...,\sum\nolimits_{f \in |\mathcal{C}|}
 {{\varepsilon _t}(f)}}_{i \in \mathcal{C}},\underbrace {0,...,0}_{i
\notin \mathcal{C}}} )$ be the distribution over all the strategies. Let ${\omega_t} = \frac{{{p_t} - u}}{{1 - \sum\nolimits_{f} {{\varepsilon _t}(f)} }}$ be the distribution
induced by AUFH-EXP3++ at the time $t$ without mixing. Then we have:
\begin{IEEEeqnarray*}{l}
\!\!\begin{array}{l}
{\mathbb{E}_{i \sim {p_s}}}{{\tilde \ell}_s}(i) = ( {1 - \sum\nolimits_{f} {{\varepsilon _s}(f)} } ){\mathbb{E}_{i \sim {\omega_s}}}{{\tilde \ell}_s}(i) + {\varepsilon _s}(i){\mathbb{E}_{i \sim u}}{{\tilde \ell}_s}(i)\\
\quad\quad \quad \quad \ \  = ( {1 - \sum\nolimits_{f} {{\varepsilon _s}(f)} } )(\frac{1}{{{\eta_s}}}\ln {\mathbb{E}_{i \sim {\omega_s}}}\exp ( - {\eta_s}({{\tilde \ell}_s}(i) \\
\quad\quad \quad \quad \quad \ - {\mathbb{E}_{j \sim {\omega_s}}} \tilde \ell_t(j))))\\
 \quad\quad \quad \quad \quad \ - \frac{( {1 - \sum\nolimits_{f} {{\varepsilon _s}(f)} } )}{{{\eta_s}}}\ln {\mathbb{E}_{i \sim {\omega_s}}}\exp ( - {\eta_s}{{\tilde \ell}_s}(i))) \\
\quad\quad \quad \quad \quad \  + {\mathbb{E}_{i \sim u}}{{\tilde \ell}_t}(i).
\end{array}\IEEEyesnumber \label{eq:AppenA}
\end{IEEEeqnarray*}

Recall that for all the strategies, we have distribution
$\omega_t = (\omega_t(1),...,\omega_t(N))$ with
\begin{IEEEeqnarray*}{l}
{\omega _t}(i) = \frac{{\exp ( - \eta_t {{\tilde L}_{t - 1}}(i))}}{{\sum\nolimits_{j = 1}^N {\exp ( - \eta_t {{\tilde L}_{t - 1}}(j))} }},
\IEEEyesnumber \label{eq:Apart2}
\end{IEEEeqnarray*}
 and  for all the channels, we have distribution  $\omega_{t,f} = (\omega_{t,f} (1),...,\omega_{t,f} (n))$
\begin{IEEEeqnarray*}{l}
{\omega_{t,f}}(f') = \frac{{\sum\nolimits_{i:f' \in i}\exp ( - \eta_t {{\tilde L}_{t - 1}}(i))}}{{\sum\nolimits_{j = 1}^N {\exp ( - \eta_t {{\tilde L}_{t - 1}}(j))} }}.\IEEEyesnumber \label{eq:Apsart2}
\end{IEEEeqnarray*}

In the second step, we use the inequalities $lnx \le x-1$ and $exp(-x) -1 +x \le x^2/2$, for all $x \ge 0$, to obtain:
\begin{IEEEeqnarray*}{l}
\begin{array}{l}
\ln {\mathbb{E}_{i \sim {\omega_s}}}\exp ( - {\eta_s}({{\tilde \ell}_s}(i) - {\mathbb{E}_{j \sim {\omega_s}}}{{\tilde \ell}_s}(j)))\\
\quad\quad \quad = \ln {\mathbb{E}_{i \sim {\omega_s}}}\exp ( - {\eta_s}{{\tilde \ell}_s}(i)) + {\eta_s}{\mathbb{E}_{j \sim {\omega_s}}}{{\tilde \ell}_s}(j)\\
\quad\quad \quad \le {\mathbb{E}_{i \sim {\omega_s}}}( {\exp ( - {\eta_s}{{\tilde \ell}_s}(i)) - 1 + {\eta_s}{{\tilde \ell}_s}(j)} )\\
\quad\quad \quad \le {\mathbb{E}_{i \sim {\omega_s}}}\frac{{\eta_s^2{{\tilde \ell}_s}{{(i)}^2}}}{2}.
\end{array}\IEEEyesnumber \label{eq:Apart1}
\end{IEEEeqnarray*}
Moreover, take expectations over all random strategies of losses ${{\tilde \ell}_s}{(i)^2}$, we have
\begin{IEEEeqnarray*}{l}
\begin{array}{l}
{\mathbb{E}_t}\left[{\mathbb{E}_{i \sim {\omega_s}}}{{\tilde \ell}_s}{(i)^2} \right] =
{\mathbb{E}_t}\left[\sum\limits_{i = 1}^N {{\omega_s}(i){{\tilde \ell}_s}{{(i)}^2}} \right]\\
= {\mathbb{E}_t}\!\!\left[\sum\limits_{i = 1}^N {{\omega_s}(i){{(\sum\limits_{f \in i} {{{\tilde \ell}_s}(f)} )}^2}}  \right]
 \le {\mathbb{E}_t}\!\!\left[ \sum\limits_{i = 1}^N {{\omega_s}(i){k_r}\!\!\sum\limits_{f \in i} {{{\tilde \ell}_s}{{(f)}^2}} }\right]
 \\= \!{\mathbb{E}_t}k_r \left[\! \sum\limits_{f = 1}^n {{{\tilde \ell}_s}{{(f)}^2}} \!\!\!\!\sum\limits_{i \in {S_r}:f \in i} \!\!\!{{\omega_s}(i)} \right]\! = \!  {k_r}{\mathbb{E}_t}\!\! \left[\!\sum\limits_{f' = 1}^n \!{{{\tilde \ell}_s}{{(f')}^2}{\omega_{s,f}}(f')} \right]\\
 \end{array}
\end{IEEEeqnarray*}
\begin{IEEEeqnarray*}{l}
\begin{array}{l}
 = {k_r}{\mathbb{E}_t}\left[ {\sum\limits_{f' = 1}^n {{{\left( {\frac{{{l_t}(f')}}{{{q_s}(f')}}{\mathds{1}_t}(f')} \right)}^2}} {\omega _{s,f}}(f')} \right]\\
 \le {k_r}{\mathbb{E}_t}\left[ {\sum\limits_{f' = 1}^n {\frac{{{\omega _{s,f}}(f')}}{{{q_s}{{(f')}^2}}}{\mathds{1}_t}(f')} } \right] = {k_r}\sum\limits_{f' = 1}^n {\frac{{{\omega _{s,f}}(f')}}{{{q_s}(f')}}} \\
 = {k_r}\sum\limits_{f' = 1}^n {\frac{{{\omega _{s,f}}(f')}}{{\left( {1 - \sum\nolimits_{f} {{\varepsilon _t}(f)} } \right)
{\omega _{s,f}}(f')
 + {\sum\nolimits_{f \in i} {{\varepsilon _t}(f)} } \left| {\left\{ {i \in \mathcal{C}:f \in i} \right\}} \right|}}}
 \le 2k_r n,
\end{array}\IEEEyesnumber \label{eq:Apart1s}
\end{IEEEeqnarray*}
where the last inequality follows the fact that $( {1 - \sum\nolimits_{f} {{\varepsilon _t}(f)} }) \ge \frac{1}{2}$ by the definition of
${{\varepsilon _t}(f)}$.

In the third step, note that ${{\tilde L}_0}(i) = 0$. Let ${\Phi _t}(\eta ) = \frac{1}{\eta }\ln \frac{1}{N}\sum\nolimits_{i = 1}^N {\exp ( - \eta {{\tilde L}_t}(i))}$ and ${\Phi _0}(\eta )=0 $. The second term in (\ref{eq:AppenA}) can be bounded by using the same technique in \cite{Bubeck12} (page 26-28).  Let us substitute  inequality (\ref{eq:Apart1s}) into (\ref{eq:Apart1}), and then substitute (\ref{eq:Apart1}) into equation  (\ref{eq:AppenA}) and sum over $t$ and take expectation over all random strategies of losses up to time $t$, we obtain
\begin{IEEEeqnarray*}{l}
\begin{array}{l}
\hspace{-.3cm}{\mathbb{E}_t}\left[ \sum\limits_{s = 1}^t {\mathbb{E}_{i \sim {p_s}}}{{\tilde \ell}_s}(i) \right]
 \le k_r n \!\sum\limits_{s = 1}^t \eta_s + \frac{{\ln N}}{\eta_t } + \!\! \sum\limits_{s = 1}^t  \!{\mathbb{E}_{i \sim u}}{{\tilde \ell}_s}(i) \\
\hspace{1.6cm} + {\mathbb{E}_t}\left[ \sum\limits_{s = 1}^{t - 1} {{\Phi _s}({\eta _{s + 1}}) - {\Phi _s}({\eta _s})}\right] + \sum\limits_{s = 1}^t {{\mathbb{E}_{{I_s} \sim {p_s}}}{\tilde{\ell}_s}(i)} .
\end{array}
\end{IEEEeqnarray*}

Then, we get
\begin{IEEEeqnarray*}{l}
R(t)  =\mathbb{E}_t \sum\limits_{s = 1}^t {{\mathbb{E}_{i \sim {p_s}}}{\tilde{\ell}_s}(i)}  - \mathbb{E}_t \sum\limits_{s = 1}^t {{\mathbb{E}_{{I_s} \sim {p_s}}}{\tilde{\ell}_s}(i)}\\
\hspace{.7cm}\le k_r n \!\sum\limits_{s = 1}^t \eta_s + \frac{{\ln N}}{\eta_t }
+  \sum\limits_{s = 1}^t  \!{\mathbb{E}_{i \sim u}}{{\tilde \ell}_s}(i)\\
\end{IEEEeqnarray*}
\begin{IEEEeqnarray*}{l}
\hspace{.7cm}\mathop  \le \limits^{(a)} k_r n \!\sum\limits_{s = 1}^t \eta_s + \frac{{\ln N}}{\eta_t }
+ k_r \sum\limits_{s = 1}^t {\sum\limits_{f = 1}^n {{\varepsilon _s}(f)} } \\
\hspace{.7cm}\mathop  \le \limits^{(b)} 2k_r n \!\sum\limits_{s = 1}^t \eta_s + \frac{{\ln N}}{\eta_t } \\
\hspace{.7cm}\mathop  \le \limits^{(c)} 2k_r n \!\sum\limits_{s = 1}^t \eta_s + k_r\frac{{\ln n}}{\eta_t }.
\end{IEEEeqnarray*}
Note that, the inequality $(a)$ holds by setting ${{\tilde \ell}_s}(i)=k_r, \forall i, s$, and the
upper bound is $k_r \sum\nolimits_{i \in C} {{\sum\nolimits_{f \in i} {{\varepsilon _t}(f)} }}=
k_r \sum\nolimits_{s = 1}^t {\sum\nolimits_{f = 1}^n {{\varepsilon _s}(f)} }$. The inequality $(b)$ holds is because of,  for every time slot $t$,
 $\eta_t \ge {\varepsilon _t}(f)$. The
 inequality $(c)$ is due to the fact that $N \le n^{k_r}$. Setting $\eta_t=\beta_t$, we prove the theorem.
  \end{IEEEproof}

 \emph{Proof of Theorem 2.}
  \begin{IEEEproof}
To defend against the m-memory-bounded adaptive adversary, we need to adopt the idea of the mini-batch protocol proposed in \cite{Arora12}.
We define a new algorithm by wrapping AUFH-EXP3++ with a mini-batching loop \cite{Dekel11}. We specify a batch
size $\tau$ and name the new algorithm AUFH-EXP3++$_\tau$. The idea is to group the overall
time slots $1,...,t$ into consecutive and disjoint mini-batches of size $\tau$. Viewing one signal mini-batch
as a round (time slot), we can use the average loss suffered during that mini-batch to feed the original AUFH-EXP3++. Note that our
new algorithm does not need to know $m$, which only appears as a constant as shown in Theorem 2. So our new
AUFH-EXP3++$_\tau$ algorithm still runs in an adaptive way without any prior about the
environment. If we set the batch $\tau= {(4{k_r}\sqrt {n\ln n} )^{ - \frac{1}{3}}}{t^{^{\frac{1}{3}}}}$ in Theorem 2 of \cite{Arora12},
we can get the regret upper bound in our Theorem 2.
  \end{IEEEproof}

\subsection{The Stochastic Regime}
 Our proofs are based on the following form of Bernstein's inequality with minor improvement as shown in \cite{Seldin14}.

\textbf{Lemma 8.} (Bernstein's inequality for martingales). Let $X_1,...,X_m$ be martingale difference sequence with
respect to filtration $\mathcal{F}=(\mathcal{F}_i)_{1 \le k \le m}$ and let $Y_k = \sum\nolimits_{j = 1}^k {{X_j}}$ be the
associated martingale. Assume that there exist positive numbers $\nu$ and $c$, such that $X_j \le c$ for all $j$ with probability
$1$ and $\sum\nolimits_{k = 1}^m {\mathbb{E}\left[ {{{\left( {{X_k}} \right)}^2}|{\mathcal{F}_{k - 1}}} \right]}  \le \nu$ with probability 1.
\begin{IEEEeqnarray*}{l}
\mathbb{P}[{Y_m} > \sqrt {2\nu b}  + \frac{{cb}}{3}] \le {e^{ - b}}.
\end{IEEEeqnarray*}

We also need to use the following technical lemma, where the proof can be found in \cite{Seldin14}.

\textbf{Lemma 9. } For any $c > 0$, we have $\sum\nolimits_{t = 0}^\infty  {{e^{ - c\sqrt t }}}  = O\left( {\frac{2}{{{c^2}}}} \right)$.

To obtain the tight regret performance for AUFH-EXP3++, we need to study and estimate the
number of times each of channel is selected up to time $t$, i.e., $N_t(f)$. We summarize it in the following lemma.

\textbf{Lemma 10. } Let $\left\{ {{{\underline \varepsilon }_t}(f)} \right\}_{t = 1}^\infty$ be non-increasing deterministic sequences, such that
${{\underline \varepsilon }_t}(f) \le {{ \varepsilon }_t}(f)$ with probability $1$ and ${{\underline \varepsilon }_t}(f) \le {{ \varepsilon }_t}(f^*)$
 for all $t$ and $f$. Define $\nu_t(f)= \sum\nolimits_{s = {1}}^t \frac{1}{{{k_r \underline \varepsilon  }_s}(f)} $, and define the event $\mathcal{E}^f_t$
\begin{IEEEeqnarray*}{l}
 {t\Delta (f) - ( {{{\tilde L}_t}(f^*) - {{\tilde L}_t}({f})} )} \\
\hspace{1.3cm} \le {\sqrt {2({\nu _t}(f) + {\nu _t}({f^*})){b_t}}  + \frac{{{(1/k_r+ 0.25)} {b_t}}}{{3k_r  {\underline \varepsilon _t}({f^*})}}}\hspace{.3cm}
(\mathcal{E}^f_t).
\end{IEEEeqnarray*}
Then for any positive sequence $b_1, b_2,...,$ and any $t^* \ge 2$ the number of times channel $f$ is played by AUFH-EXP3++ up to
round $t$ is bounded as:
\begin{IEEEeqnarray*}{l}
\begin{array}{l}
\mathbb{E}[{N_t}(f)] \le \left( {{t^*} - 1} \right) + \sum\limits_{s = {t^*}}^t {{e^{ - {b_s}}}}  + k_r \sum\limits_{s = {t^*}}^t {{\varepsilon _s}(f){\mathds{1}_{\{ \mathcal{E}_t^f\} }}} \\
    \hspace{1.6cm}               + \sum\limits_{s = {t^*}}^t {{e^{ - {\eta _s}{h_{s - 1}}(f)}}},
\end{array}
\end{IEEEeqnarray*}
where
\begin{IEEEeqnarray*}{l}
\begin{array}{l}
{h_t}(f) = t\Delta (f) - \sqrt {2t{b_t}\left( {\frac{1}{{{k_r}{{\underline \varepsilon  }_t}(f)}} + \frac{1}{{{k_r}{{\underline \varepsilon  }_t}({f^*})}}} \right)}  - \frac{{(\frac{1}{4} + \frac{1}{k_r}){b_t}}}{{3{{\underline \varepsilon  }_t}({f^*})}}.
\end{array}
\end{IEEEeqnarray*}

\begin{IEEEproof}
Note that the elements of the martingale difference sequence $\{ {\Delta (f)
- ({{\tilde \ell}_t}(f) - {{\tilde \ell}_t}({f^*}))} \}_{t = 1}^\infty$ by $\max \{ \Delta (f) +{{\tilde \ell}_t}({f^*}) \}= {\frac{1}{{{k_r}{{\underline \varepsilon  }_t}({f^*})}}} +1$. Since ${{{\underline \varepsilon  }_t}({f^*})} \le {{{ \varepsilon  }_t}({f^*})} \le 1/(2n) \le 1/4$, we can simplify
the upper bound by using ${\frac{1}{{{{k_r\underline \varepsilon   }_t}({f^*})}}} +1 \le
    \frac{{(\frac{1}{4} + \frac{1}{k_r})}}{{{{\underline \varepsilon   }_t}({f^*})}}$.

  We further note that
   \begin{displaymath}
\begin{array}{l}
\sum\limits_{s = 1}^t {{\mathbb{E}_s}\left[{{(\Delta (f) - ({{\tilde \ell}_s}(f) - {{\tilde \ell}_s}({f^*})))}^2}\right]} \\
 \hspace{1.1cm}  \le \sum\limits_{s = 1}^t {{\mathbb{E}_s}\left[{{({{\tilde \ell}_s}(f) - {{\tilde \ell}_s}({f^*}))}^2}\right]} \\
 \hspace{1.1cm}  = \sum\limits_{s = 1}^t {\left( {{\mathbb{E}_s}\left[({{\tilde \ell}_s}{{(f)}^2}\right] + {E_s}\left[({{\tilde \ell}_s}{{({f^*})}^2}\right]} \right)} \\
 \end{array}
 \end{displaymath}
    \begin{displaymath}
\begin{array}{l}
 \hspace{1.1cm}  \le \sum\limits_{s = 1}^t \left( {\frac{1}{{{q_s}(f)}} + \frac{1}{{{q_s}({f^*})}}} \right) \\
 \hspace{1.1cm} \mathop \le \limits^{(a)} \sum\limits_{s = 1}^t {\left( {\frac{1}{{k_r{\varepsilon  _s}(f)}} + \frac{1}{{k_r{\varepsilon  _s}({f^*})}}} \right)} \\
 \hspace{1.1cm}  \le \sum\limits_{s = 1}^t {\left( {\frac{1}{{k_r{{\underline \varepsilon   }_s}(f)}} +
 \frac{1}{{k_r{{\underline \varepsilon   }_s}({f^*})}}} \right)}  = {\nu _s}(f) + {\nu _s}({f^*})
\end{array}
\end{displaymath}
with probability $1$.  The above inequality (a) is due to the fact that $q_t(f) \ge \sum\nolimits_{f \in i} {{\varepsilon _t}(f)} \left| {\left\{ {i \in \mathcal{C}:f \in i} \right\}} \right|$. Since each $f$ only belongs to one of the covering strategies $i \in \mathcal{C}$, $\left| {\left\{ {i \in \mathcal{C}:f \in i} \right\}} \right|$ equals to 1 at time slot $t$ if channel $f$ is selected. Thus, $q_t(f) \ge \sum\nolimits_{f \in i} {{\varepsilon _t}(f)}= k_r{\varepsilon_t}(f)$.

Let $\mathcal{\bar E}_t^f$ denote the complementary of event
 $\mathcal{E}_t^f$. Then by the Bernstein's inequality $\mathbb{P}[\mathcal{\bar E}_t^f] \le e^{-b_t}$. The number of
 times the channel $f$ is selected up to round $t$ is bounded as:
\begin{IEEEeqnarray*}{l}
 \begin{array}{l}
 \mathbb{E}[{N_t}(f)] = \sum\limits_{s = 1}^t {\mathbb{P}[A_s= f]} \\
 \hspace{4em} = \sum\limits_{s = 1}^t {\mathbb{P}[A_s= f|\mathcal{E}_{s - 1}^f]P[\mathcal{E}_{s - 1}^f]} \\
  \hspace{5em}+ \mathbb{P}[A_s= f|\overline {\mathcal{E}_{s - 1}^f} ]P[\overline {\mathcal{E}_{s - 1}^f} ]\\
 \hspace{4em} \le \sum\limits_{s = 1}^t {\mathbb{P}[A_s= f|\mathcal{E}_{s - 1}^f]} {\mathds{1}_{\{ \mathcal{E}_{s - 1}^f\} }} +
  \mathbb{P}[\overline {\mathcal{E}_{s - 1}^S} ]\\
  \hspace{4em} \le \sum\limits_{s = 1}^t {\mathbb{P}[A_s= f|\mathcal{E}_{s - 1}^f]} {\mathds{1}_{\{ \mathcal{E}_{s - 1}^f\} }} + {e^{ - {b_{s - 1}}}}.
\end{array}
\end{IEEEeqnarray*}
We further upper bound $ {\mathbb{P}[A_s= f|\mathcal{E}_{s - 1}^f]} {\mathds{1}_{\{ \mathcal{E}_{s - 1}^f\} }} $ as follows:
    \begin{displaymath}
\begin{array}{l}
\mathbb{P} {[A_s= f|{\cal \mathcal{E}}_{s - 1}^f]} {\mathds{1}_{\{ {\cal \mathcal{E}}_{s - 1}^f\} }}
= {q_s}(f){\mathds{1}_{\{ {\cal \mathcal{E}}_{s - 1}^f\} }}\\
  \hspace{4em} \le ({\omega_t}(f) + k_r{\varepsilon_s}(f)){\mathds{1}_{\{ {\cal \mathcal{E}}_{s - 1}^f\} }}\\
 \hspace{4em} =({k_r\varepsilon_s}(f) + \frac{{\sum\nolimits_{i:f \in i} {{w_{s - 1}}\left( i \right)} }}{{{W_{s - 1}}}}){\mathds{1}_{\{ {\cal \mathcal{E}}_{s - 1}^f\} }}\\
    \hspace{4em} =({k_r\varepsilon  _s}(f) + \frac{{\sum\nolimits_{i:f \in i}^{} {{e^{ - {\eta _s}{ \tilde L_{s - 1}}(i)}}} }}{{\sum\nolimits_{i = 1}^N {{e^{ - {\eta _t}{ \tilde L_{s - 1}}(i)}}} }}){\mathds{1}_{\{ {\cal \mathcal{E}}_{s - 1}^f\} }}\\
  \hspace{4em}\mathop \le \limits^{(a)} ({k_r\varepsilon  _s}(f) + {e^{ - {\eta _s}\left( {{{\tilde L}_{s - 1}}(i) - {{\tilde L}_{s- 1}}({i^*})} \right)}}){\mathds{1}_{\{ {\cal \mathcal{E}}_{s - 1}^f\} }}\\
    \hspace{4em}\mathop \le \limits^{(b)} ({k_r\varepsilon  _s}(f) + {e^{ - {\eta _s}\left( {{{\tilde L}_{s - 1}}(f) - {{\tilde L}_{s - 1}}({f^*})} \right)}}){\mathds{1}_{\{ {\cal \mathcal{E}}_{s - 1}^f\} }}\\
   \hspace{4em}\mathop \le \limits^{(c)}  k_r{\varepsilon  _s}(f){\mathds{1}_{\{ {\cal \mathcal{E}}_{s - 1}^f\} }} + {e^{ - {\eta _s}{h_{s - 1}}(f)}}.
\end{array}
  \end{displaymath}
The above inequality (a) is due to the fact that channel
$f$ only belongs to one selected strategy $i$ in $t-1$, inequality (b) is because of the  cumulative
regret of each strategy is great than the cumulative regret of each channel that belongs to the
  strategy, and the last inequality (c) we used the fact that $\frac{t}{{{{\underline \varepsilon   }_t}(f)}}$ is a non-increasing
 sequence ${\upsilon _t}(f) \le \frac{t}{{{{k_r \underline \varepsilon   }_t}(f)}}$. Substitution of this result back into
 the computation of $ \mathbb{E}[{N_t}(f)]$ completes the proof. \end{IEEEproof}

\emph{Proof of Theorem 3.}
\begin{IEEEproof}
The proof is based on Lemma 10. Let $b_t =ln(t \Delta(f)^2)$ and ${{{\underline \varepsilon  }_t}(f)}
={{{ \varepsilon  }_t}(f)}$. For any $c \ge 18$ and any $t \ge t^*$, where $t^*$ is the minimal integer for which
 ${t^*} \ge \frac{{4{c^2}n \ln {{({t^*}\Delta {{(f)}^2})}^2}}}{{\Delta {{(f)}^4}\ln (n)}}$, we have
     \begin{displaymath}
\begin{array}{l}
{h_t}(f) = t\Delta (f) - \sqrt {2t{b_t}\left( {\frac{1}{{k_r{\varepsilon _t}(f)}} + \frac{1}{{k_r{\varepsilon _t}({f^*})}}} \right)}  - \frac{{\left( {\frac{1}{4} + \frac{1}{k_r}} \right){b_t}}}{{3{\varepsilon _t}({f^*})}}\\
   \hspace{2.43em} \ge t\Delta (f) - 2\sqrt {\frac{{t{b_t}}}{{k_r{\varepsilon _t}(f)}}}  - \frac{{\left( {\frac{1}{4} + \frac{1}{k_r}} \right){b_t}}}{{3{\varepsilon _t}(f)}}\\
  \hspace{2.43em} = t\Delta (f)(1 - \frac{2}{{\sqrt {k_r c} }} - \frac{{\left( {\frac{1}{4} + \frac{1}{k_r}} \right)}}{{3c}})\\
  \hspace{2.43em} \mathop  \ge \limits^{(a)} t\Delta (f)(1 - \frac{2}{{\sqrt c }} - \frac{{1.25}}{{3c}}) \ge \frac{1}{2}t\Delta (f).
\end{array}
  \end{displaymath}
The above inequality (a) is due to the fact that $(1 - \frac{2}{{\sqrt {k_r c} }} - \frac{{\left( { \frac{1}{4}+\frac{1}{k_r}} \right)}}{{3c}}$ is
an increasing function with respect to $k_r (k_r \ge 1)$. Plus, as indicated in work \cite{Seldin13}, by a bit more sophisticated  bounding $c$ can be made almost as small as 2 in our case. By substitution of the lower bound on $h_t(f)$ into Lemma 10, we have
     \begin{displaymath}
\begin{array}{l}
\!\!\!\mathbb{E}[{N_t}(f)] \le {t^*} + \frac{{\ln (t)}}{{\Delta {{(f)}^2}}} + k_r \!\! \frac{{c\ln {{(t)}^2}}}{{\Delta {{(f)}^2}}}  +  \sum\limits_{s = 1}^t \!\!\left(\!{{e^{ - \frac{{\Delta (f)}}{4}\sqrt {\frac{{(s - 1)ln(n)}}{n}} }}}\!\right)\\
 \hspace{3.5em} \le k_r\frac{{c\ln {{(t)}^2}}}{{\Delta {{(f)}^2}}} + \frac{{\ln (t)}}{{\Delta {{(f)}^2}}} + O(\frac{{{n
 }}}{{\Delta {{(f)}^2}}}) + {t^*},
\end{array}
  \end{displaymath}
  where we used lemma 3 to bound the sum of the exponents. In addition, please
  note that $t^*$ is of the order $O(\frac{{k_r n}}{{\Delta {{(f)}^4}\ln (n)}})$.
\end{IEEEproof}

\emph{Proof of Theorem 4.}
\begin{proof} The proof is based on the similar idea of Theorem 2 and Lemma 10. Note that
by our definition ${{\hat \Delta }_t}(f) \le 1$ and the sequence ${\underline \varepsilon _t}(f) = {\underline \varepsilon  _t} =
\min \{ \frac{1}{{2n}},{\beta _t},\frac{{c\ln {{(t)}^2}}}{t}\} $ satisfies the condition of Lemma 10. Note that when ${\beta _t} \ge
\frac{{c\ln {{(t)}^2}}}{t}\}$, i.e., for $t$ large enough such that
$
t \ge \frac{{4{c^2}\ln {{(t)}^4}n }}{{\ln (n)}}
$, we have ${\underline \varepsilon  _t}=\frac{{c\ln {{(t)}^2}}}{t}$. Let $b_t=ln(t)$ and let $t^*$ be
large enough, so that for all $t \ge t^*$ we have $t \ge \frac{{4{c^2}\ln {{(t)}^4}n }}{{\ln (n)}}$ and
$t \ge e^{\frac{1}{\Delta(f)^2}}$. With these parameters and conditions on hand, we are going to bound the
rest of the three terms in the bound on $\mathbb{E}[N_t(f)]$ in Lemma 10. The upper bound of
 $\sum\nolimits_{s = {t^*}}^t {{e^{ - {b_s}}}} $ is easy to obtain. For bounding
 $k_r\sum\nolimits_{s = {t^*}}^t {{\varepsilon _s}(f){\mathds{1}_{\{ \mathcal{E}_{s - 1}^f\} }}}$,  we note that $\mathcal{E}_{t}^f$ holds and we have
      \begin{displaymath}
\begin{array}{l}
{{\hat \Delta }_t}(f) \ge \frac{1}{t}(\mathop {\max }\limits_k ({{\tilde L}_t}(k)) - {{\tilde L}_t}(f)) \ge \frac{1}{t}({{\tilde L}_t}({f^*}) - {{\tilde L}_t}(f))\\
 \hspace{2.43em} \ge \frac{1}{t}{h_t}(f) = \frac{1}{t}\left( {t\Delta (f) - 2\sqrt {\frac{{t{b_t}}}{{k_r{{\underline \varepsilon  }_t}}}}  - \frac{{(\frac{1}{4} + \frac{1}{k_r}){b_t}}}{{3{{\underline \varepsilon  }_t}}}} \right)\\
 \hspace{2.43em} = \frac{1}{t}\left( {t\Delta (f) - \frac{{2t}}{{\sqrt {c k_r \ln (t)} }} - \frac{{(\frac{1}{4} + \frac{1}{k_r})t}}{{3c\ln (t)}}} \right)\\
 \hspace{2.43em} \mathop  \ge \limits^{(a)} \frac{1}{t}\left( {t\Delta (f) - \frac{{2t}}{{\sqrt {c\ln (t)} }} - \frac{{1.25t}}{{3c\ln (t)}}} \right)\\
\hspace{2.43em} \mathop  \ge \limits^{(b)} \Delta (f)\left( {1 - \frac{2}{{\sqrt c }} - \frac{{1.25}}{{3c}}} \right) \ge \frac{1}{2}\Delta (f),
\end{array}
  \end{displaymath}
  where the inequality (a) is due to the fact that $\frac{1}{t}( t\Delta (f) - \frac{2t}{\sqrt {c k_r \ln (t)} } -
  \frac{(\frac{1}{4} + \frac{1}{k_r})t}{3c\ln (t)} )$ is
an increasing function with respect to $k_r (k_r \ge 1)$ and the inequality (b) due to the fact that for $t \ge t^*$ we have $\sqrt {ln(t)}
\ge 1/\Delta(f).$ Thus,
\begin{displaymath}
{\varepsilon _n}(f){\mathds{1}_{\{ \mathcal{E}_{n - 1}^f\} }}
\le \frac{{c{{\left( {\ln t} \right)}^2}}}{{t{{\hat \Delta }_t}{{(f)}^2}}} \le
 \frac{{4{c^2}{{\left( {\ln t} \right)}^2}}}{{t\Delta {{(f)}^2}}}
 \end{displaymath}
 and $k_r\sum\nolimits_{s = {t^*}}^t {{\varepsilon _s}(f){\mathds{1}_{\{ \mathcal{E}_{n - 1}^f\} }}} = O\left(
 {\frac{{k_r\ln {{\left( t \right)}^3}}}{{\Delta {{(f)}^2}}}} \right)$. Finally, for the last term in Lemma 10, we have
 already get $h_t(f) \ge \frac{1}{2}\Delta(f)$ for $t \ge t^*$ as an intermediate step in the calculation of bound
 on ${{{\hat \Delta }_t}(f)}$. Therefore, the last term
 is bounded in a order of $O(\frac{{{n
 }}}{{\Delta {{(f)}^2}}})$. Use all these results together we obtain the results of the theorem. Note that the
 results holds for any $\eta_t \ge \beta_t$.
 \end{proof}

\subsection{Mixed Adversarial and Stochastic Regime}
\emph{Proof of Theorem 5.}
\begin{proof}
The proof of the regret performance in the mixed adversarial and stochastic regime is simply a combination of the performance of
the AUFH-EXP3++$^\emph{AVG}$ algorithm in adversarial and stochastic regimes. It is very straightforward from Theorem 1 and Theorem
3.
 \end{proof}
\emph{Proof of Theorem 6.}
\begin{proof}
Similar as above, the proof is very straightforward from Theorem 2 and Theorem 3.
 \end{proof}

\subsection{Contaminated Stochastic Regime}
\emph{Proof of Theorem 7.}
\begin{proof}
The key idea of proving the regret bound under  moderately contaminated stochastic  regime
 relies on how to estimate the performance loss by taking into account the contaminated pairs. Let $\mathds{1}
 _{t,f}^\star$ denote the indicator functions of the occurrence of contamination at location $(t,f)$, i.e.,
 $\mathds{1} _{t,f}^\star$  takes value $1$ if contamination occurs and $0$ otherwise.
 Let $m_t(f)= \mathds{1}  _{t,f}^\star \tilde \ell_t(f) + (1-\mathds{1}  _{t,f}^\star)\mu(f)$.  If either base arm $f$
 was contaminated on round $t$ then $m_t(f)$ is adversarially assigned a value of loss that is
 arbitrarily affected by some adversary, otherwise we use the expected loss. Let  ${M_t}(f) = \sum\nolimits_{s = 1}^t {{m_t}(f)}$
 then $\left( {{M_t}({f}) - {M_t}(f^*)} \right) - \left( {{{\tilde L}_t}({f}) - {{\tilde L}_t}(f^*)} \right)$ is a martingale.
 After
 $\tau$ steps, for $t \ge \tau$,
  \begin{displaymath}
\begin{array}{l}
\left( {{M_t}({f}) - {M_t}(f^*)} \right) \ge t\min \{ \mathds{1}  _{t,f}^\star ,\mathds{1}  _{t,f^*}^\star \} ({\tilde \ell_t}(f) - {\tilde \ell_t}({f^*}))\\
\hspace{6em} + t\min \{ 1 - \mathds{1}  _{t,f}^\star ,1 - \mathds{1}  _{t,f^*}^\star \} (\mu ({f}) - \mu (f^*))\\
 \hspace{3.6em} \ge  - \zeta t\Delta (f) + (t - \zeta t\Delta (f))\Delta (f) \ge (1-2\zeta){t\Delta (f)}.
\end{array}
  \end{displaymath}

Define the event $\mathcal{Z}_t^f$:
  \begin{displaymath}
(1-2\zeta)t\Delta (f) - \left( {{{\tilde L}_t}({f}) - {{\tilde L}_t}(f^*)} \right)
\le 2\sqrt {{\nu _t}{b_t}}  + \frac{{\left( {\frac{1}{4} + \frac{1}{k_r}} \right){b_t}}}{{3{{\underline \varepsilon  }_t}}},
  \end{displaymath}
where ${\underline \varepsilon  }_t$ is defined in the proof of Theorem 3 and $\nu _t = \sum\nolimits_{s = 1}^t
 {\frac{1}{{{{k_r\underline \varepsilon  }_t}}}}$. Then by Bernstein's inequality
  $\mathbb{P}[\mathcal{Z}_t^f] \le e^{-b_t}$. The remanning proof is identical to the proof of Theorem 3.

  For the regret performance in the moderately contaminated stochastic regime, according to our definition with the attacking strength
  $\zeta \in [0,1/4]$, we only need to replace  $\Delta(f)$  by $\Delta(f)/2$ in Theorem 5.
\end{proof}

\section{The Computational Efficient Implementation of the AUFH-EXP3++ Algorithm}
The implementation of algorithm $1$ requires the computation of probability distributions and storage of $N$ strategies, which is
obvious to have a time and space complexity $O(n^{k_r})$. As the number of channels increases, the strategy will become
exponentially large, which is very hard to be scalable and results in low efficiency. To address this important problem,
we propose a computational efficient enhanced algorithm by utilizing the dynamic programming techniques, as shown in Algorithm 2. The key idea
of the enhanced algorithm is to select the receiving channels one by one until $k_r$ channels are chosen, instead of choosing a strategy
from the large strategy space in each time slot.

We use $S\left( {\bar f,\bar k} \right)$ to denote the strategy set of which each strategy selects $\bar k$ channels from $\bar f, \bar f+1,
\bar f, ..., n$. We also use $\bar S\left( {\bar f,\bar k} \right)$ to denote the strategy set of which each strategy selects $\bar k$ channels from
channel $1,2,..., \bar f$. We define $
{W_t}(\bar f,\bar k) = \sum\nolimits_{i \in S(\bar f,\bar k)} {\prod\nolimits_{f \in i} {{w_t}(f)} }$ and $
{W_t}(\bar f,\bar k) = \sum\nolimits_{i \in \bar S(\bar f,\bar k)} {\prod\nolimits_{f \in i} {{w_t}(f)} },
$
Note that they have the following properties:
\begin{IEEEeqnarray*}{l}
{W_t}(\bar f,\bar k) = {W_t}(\bar f + 1,\bar k) + {w_t}(\bar f){W_t}(\bar f + 1,\bar k - 1),
\IEEEyesnumber \label{eq:Bpart1s}
\end{IEEEeqnarray*}
\begin{IEEEeqnarray*}{l}
{W_t}(\bar f,\bar k) = {W_t}(\bar f - 1,\bar k) + {w_t}(\bar f){W_t}(\bar f - 1,\bar k - 1),
\IEEEyesnumber \label{eq:Bpart2s}
\end{IEEEeqnarray*}
which implies both ${W_t}(\bar f,\bar k)$ and ${\bar W_t}(\bar f,\bar k)$ can be calculated in $O(k_rn)$ (Letting ${W_t}(\bar f,0)=1$ and
$W(n + 1,\bar k) = \bar W(0,\bar k) = 0$) by using dynamic programming for all $1 \le \bar f \le n$ and $1 \le \bar k \le k_r$.

In step 1, a strategy should be drawn from $\binom{n}{k_r}$ strategies. Instead of drawing a strategy, we select channel for the strategy
one by one until a strategy is found. Here, we select channels one by one in the increasing order of channel indices, i.e., we determine
whether the channel $1$ should be selected, and the channel $2$, and so on. For any channel $f$, if $k \le k_r$ channels have been chosen in channel
$1,..,f-1$, we select channel $f$ with probability
\begin{IEEEeqnarray*}{l}
\frac{{{w_{t - 1}}(f){W_t}(f + 1,{k_r} - k - 1)}}{{{W_{t - 1}}(f,{k_r} - k)}}\IEEEyesnumber \label{eq:Bpart3s}
\end{IEEEeqnarray*}
and not select $f$ with probability $
\frac{{{W_t}(f + 1,{k_r} - k - 1)}}{{{W_{t - 1}}(f,{k_r} - k)}}.
$
Let $w(f) = {w_{t - 1}}(f)$ if channel $f$ is selected in the strategy $i$; $w(f) = 0$ otherwise. Obviously, $w(f)$ is actually the weight
of $f$ in the strategy weight. In our algorithm, ${w_{t - 1}}(f) = \prod\nolimits_{f = 1}^n {w(f)}$. Let $c(f)=1$ if $f$ is selected in $i$;
$c(f)=0$ otherwise. The term $\sum\nolimits_{f = 1}^{\bar f} {c(f)}$ denotes the number of channels chosen among channel $1,2,...,\bar f$ in strategy
$i$. In this implementation, the probability that a strategy $i$ is selected is
$
\prod\limits_{\bar f = 1}^n {\frac{{w(\bar f){W_{t - 1}}(\bar f + 1,{k_r} - \sum\nolimits_{f = 1}^{\bar f} {c\left( f \right)} )}}{{{W_{t - 1}}(\bar f,{k_r} - \sum\nolimits_{f = 1}^{\bar f - 1} {c\left( f \right)} )}}}  = \frac{{\prod\limits_{\bar f = 1}^n {w(\bar f)} }}{{{W_{t - 1}}(1,{k_r})}}
 = \frac{{{w_{t - 1}}(i)}}{{{W_{t - 1}}}}.
$
This probability is equivalent to that in Algorithm 1, which implies the implementation is correct.
Because we do not maintain $w_t(i)$, it is impossible to compute $q_t(f)$ as we have described in Algorithm 1. Then
$q_t(f)$ can be computed within $O(nk_r)$ as in Eq.(4) for each round.

Moreover, for the exploration parameters  $\varepsilon_t(f)$, since there are $k_r$ parameters of $\varepsilon_t(f)$ in the
last term of Eqs. (\ref{eq:Bpart4s}) and there are $n$ channels, the storage complexity is $O(k_rn)$. Similarly, we have the time complexity
$O(k_rnt)$ for the maintenance of exploration parameters  $\varepsilon_t(f)$. Based on the above analysis, we can summarize the conclusions into the following theorem.

\textbf{Theorem 11.} The Algorithm 2 has time complexity $O(k_rnt)$ and space complexity $O(k_rn)$, which has the linear scalability
along with rounds $t$, and parameters $k_r$ and $n$.

\newcounter{mytempeqncnt}
\begin{figure*}[!t]
\normalsize
\setcounter{mytempeqncnt}{\value{equation}}
\setcounter{equation}{5}
\begin{equation}
(1 - \sum\nolimits_{f = 1}^n {{\varepsilon _t}(f)} )\frac{{\sum\nolimits_{k = 0}^{{k_r} - 1} {{{\bar W}_{t - 1}}(f - 1,k){w_{t - 1}}(f){W_{t - 1}}(f + 1,{k_r} - k - 1)} }}{{{W_{t - 1}}(1,k)}}\\ + \sum\limits_{f \in i} {\varepsilon _t}(f)\left| {i \in C:f \in i} \right|
\IEEEyesnumber \label{eq:Bpart4s}
\end{equation}
\setcounter{equation}{\value{mytempeqncnt}}
\hrulefill
\vspace*{4pt}
\end{figure*}

\begin{algorithm}
\caption{An Computational Efficient Implementation of AUFH-EXP3++}
\begin{algorithmic}
\STATE \textbf{Input}: $n, k_r, t$,  and See text for definition of $\eta_t$ and $\xi_t(f)$.
\STATE \textbf{Initialization}: Set initial channel weight $w_0(f)=1, \forall f \in [1,n]$. Let $W_t(f,0)=1$ and
$W(n+1,k)= \bar W(0,k)=0$ and compute $W_0(f,k)$ and $\bar W_0(f,k)$ follows Eqs. (\ref{eq:Bpart1s}) and (\ref{eq:Bpart2s}), respectively.
\FOR {time slot $t=1,2,...$}
\STATE 1: The receiver selects a channel $f, \forall f \in [1,n]$ one by one according to the channel's probability distribution computed following
 Eq. (\ref{eq:Bpart3s}) until a strategy with $k_r$ chosen channels are selected.

\STATE 2: The receiver computes the probability $q_t(f),\forall f \in \left[ {1,n} \right]$ according to Eq. (\ref{eq:Bpart4s}).

\STATE 3: The receiver calculates the loss for channel $f$, $\ell_{t-1}(f), \forall f \in I_t$ based on the received
channel gain $g_{t-1}(f)$ by using $\ell_{t-1}(f) = 1- g_{t-1}(f)$. Compute the estimated loss $\tilde{\ell}_{t}(f), \forall f \in [1,n]$ as follows:
\begin{IEEEeqnarray*}{l}
{\tilde{\ell}_t}(f) = \left\{ \begin{array}{l}
\frac{{{\ell_t}(f)}}{{{q_t}(f)}} \quad  \text{\emph{if channel}}  f \in {I_t} \\
0         \ \ \quad\quad \emph{otherwise}.
\end{array} \right.
\end{IEEEeqnarray*}

\STATE 4: The receiver updates all channel weights as ${w_t}\left( f \right) = {w_{t - 1}}\left( f \right){e^{ - \eta_t { {\tilde \ell}_t}(f)}} = {e^{ - \eta_t {{\tilde L}_t}(f)}}, \forall f \in [1,n]$, and computes $W_t(f,k)$ and $\bar W_t(f,k)$ follows Eqs. (\ref{eq:Bpart1s}) and (\ref{eq:Bpart2s}), respectively.
\ENDFOR
\end{algorithmic}
\end{algorithm}

Besides, because of the channel selection probability for $q_t(f)$ and the updated weights of Algorithm 2 equals to
Algorithm 1, all the performance results in Section IV still hold for Algorithm 2.

\begin{figure*}
\includegraphics[width=2.3in]{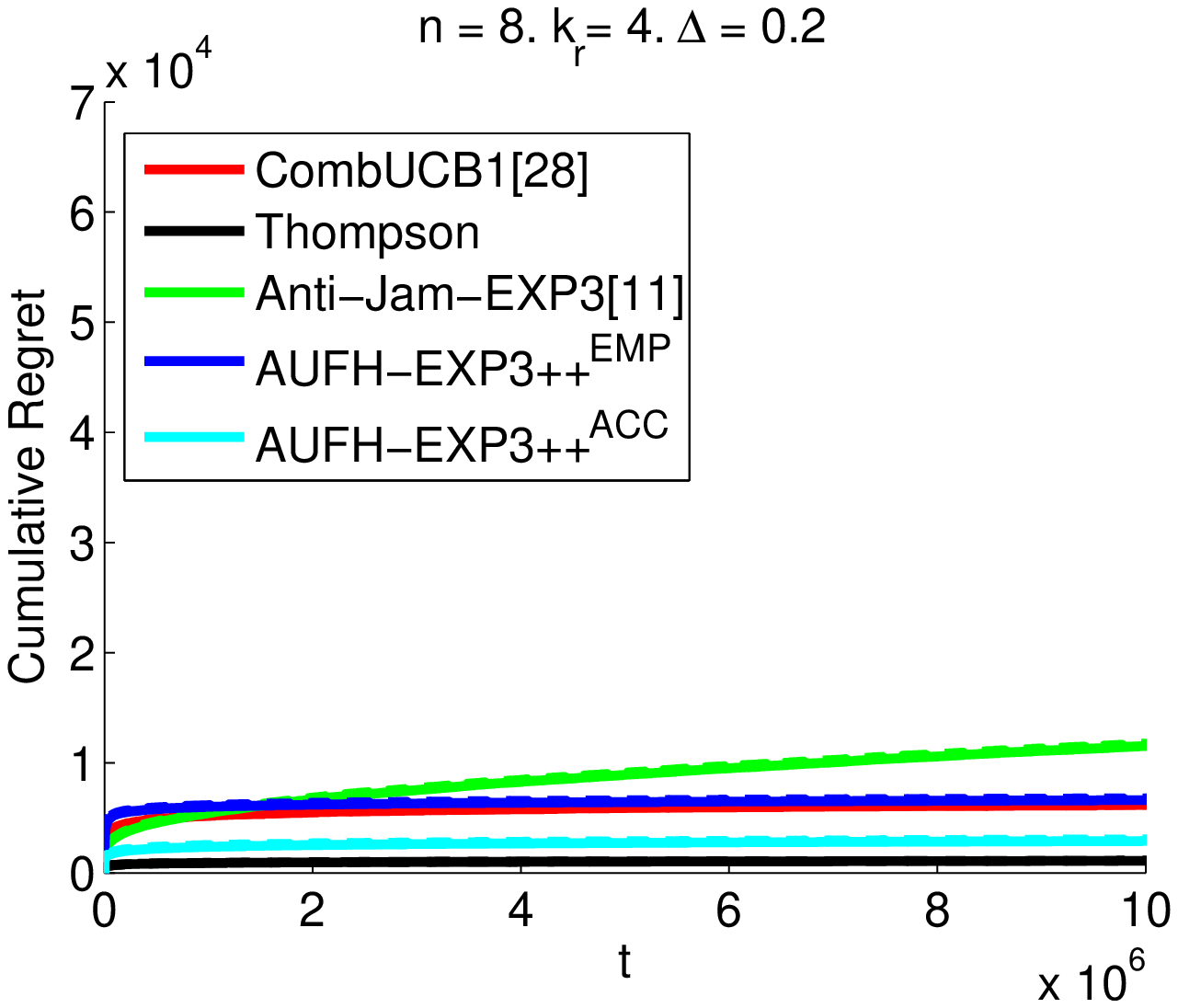}
\includegraphics[width=2.3in]{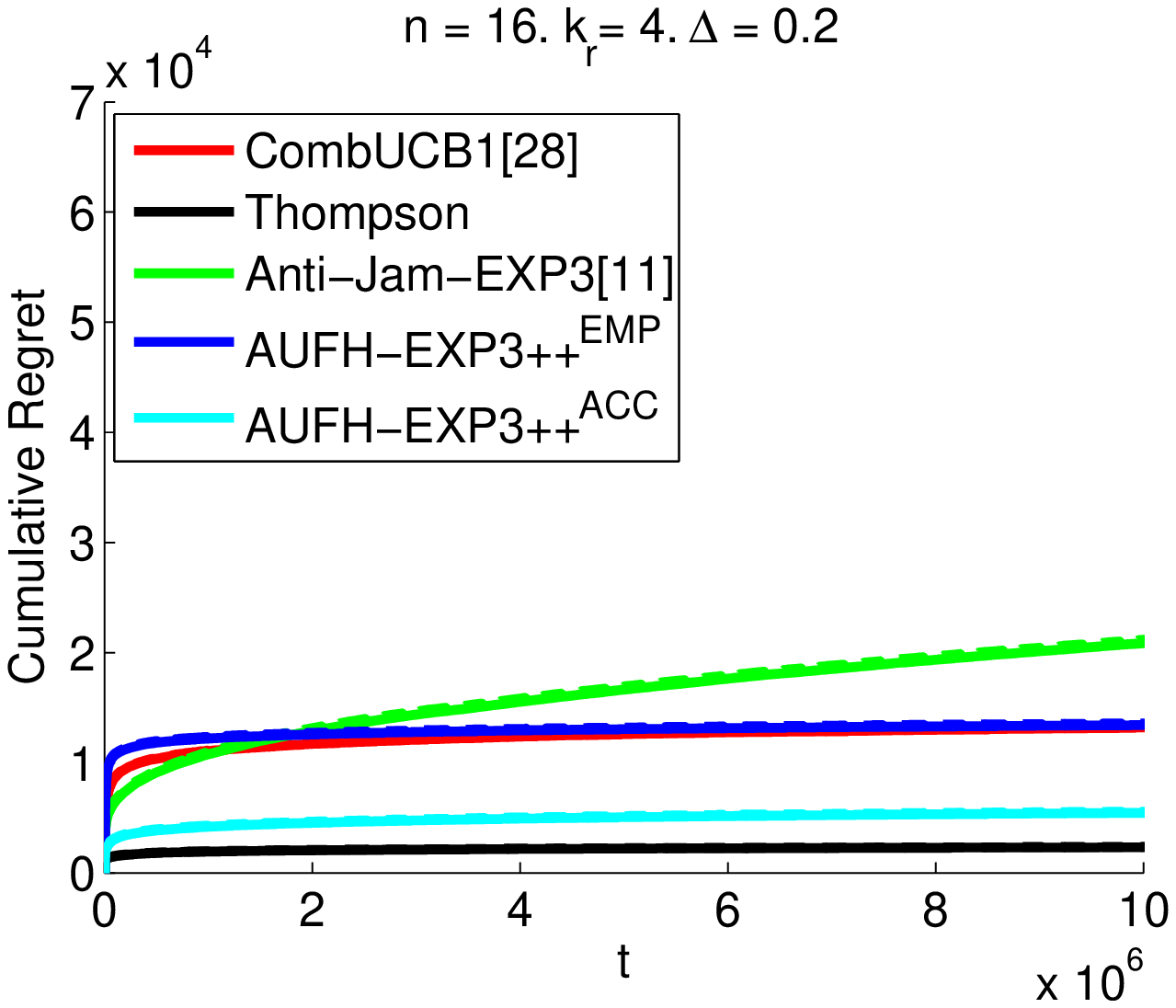}
\includegraphics[width=2.3in]{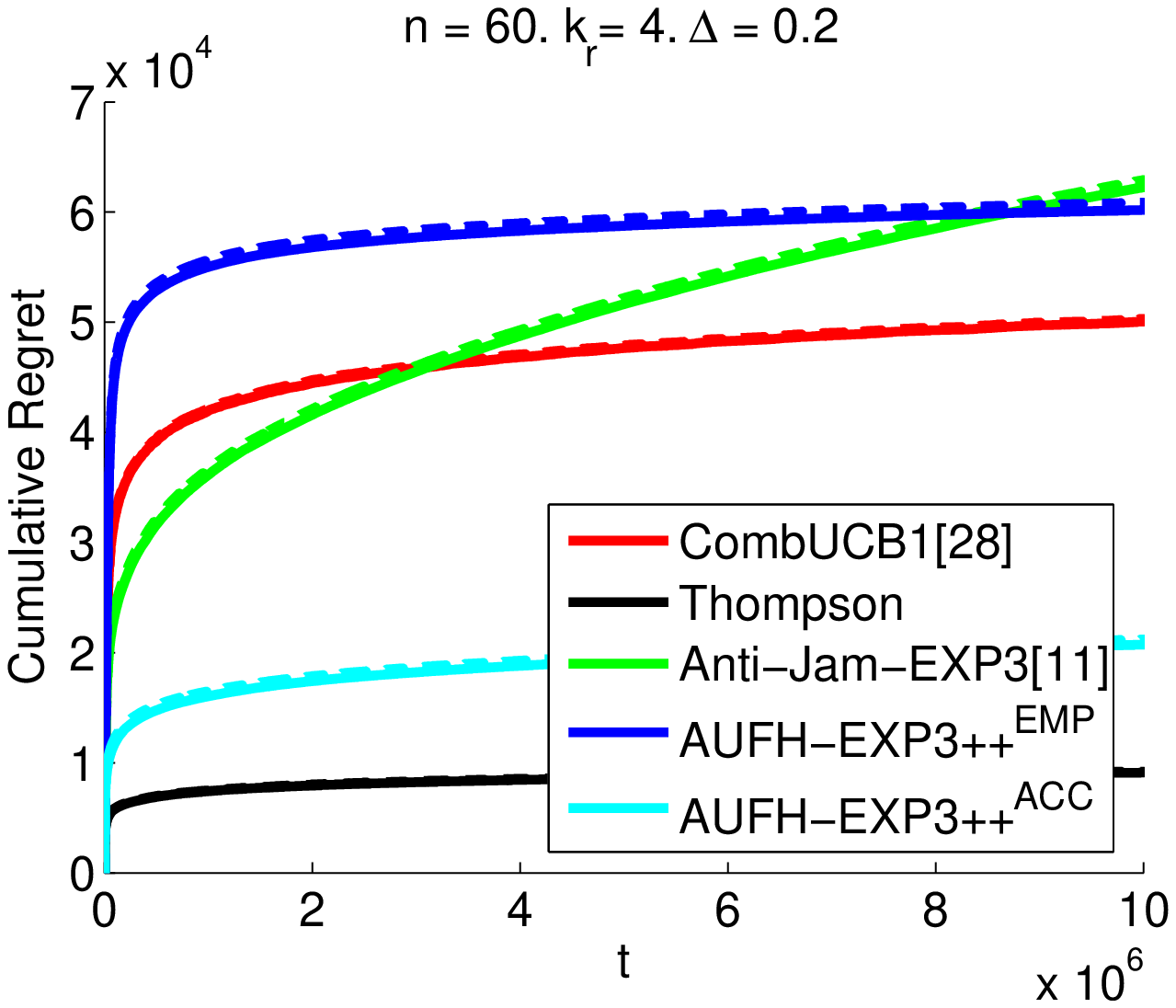}
  \caption{Performance Comparison in the Stochastic Regime.}
\end{figure*}
%

\begin{figure*}
\includegraphics[width=2.33in]{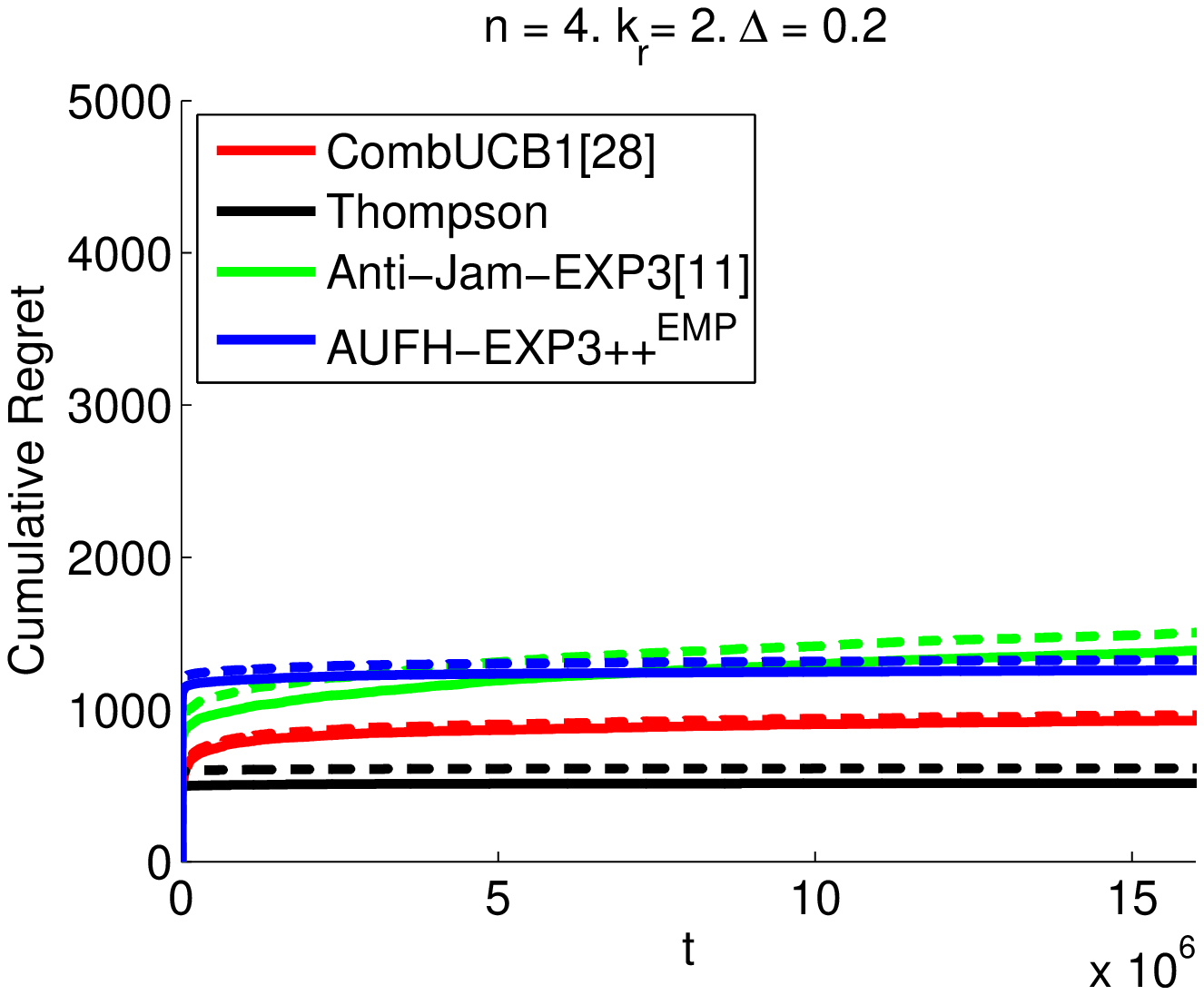}
\includegraphics[width=2.33in]{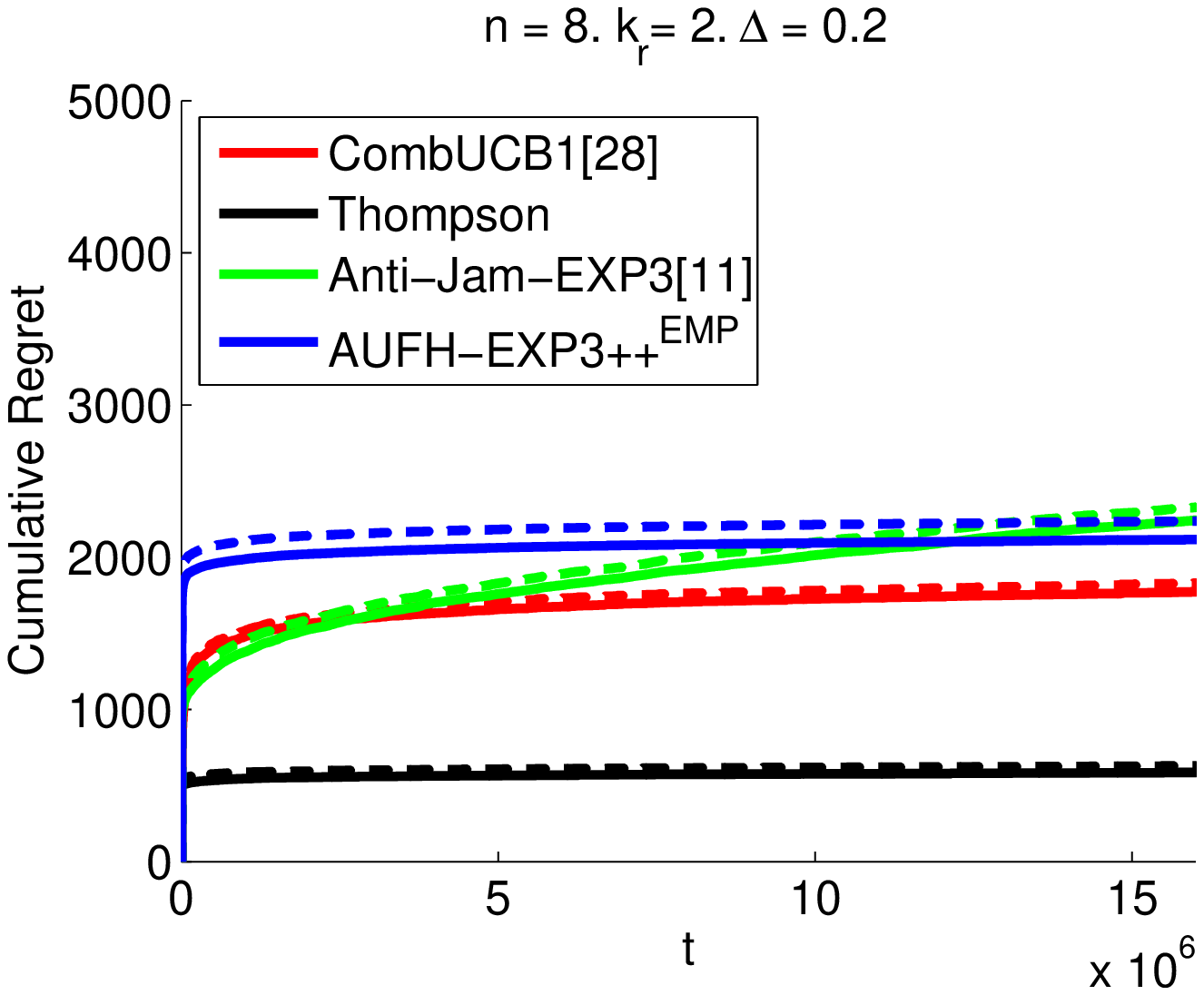}
\includegraphics[width=2.33in]{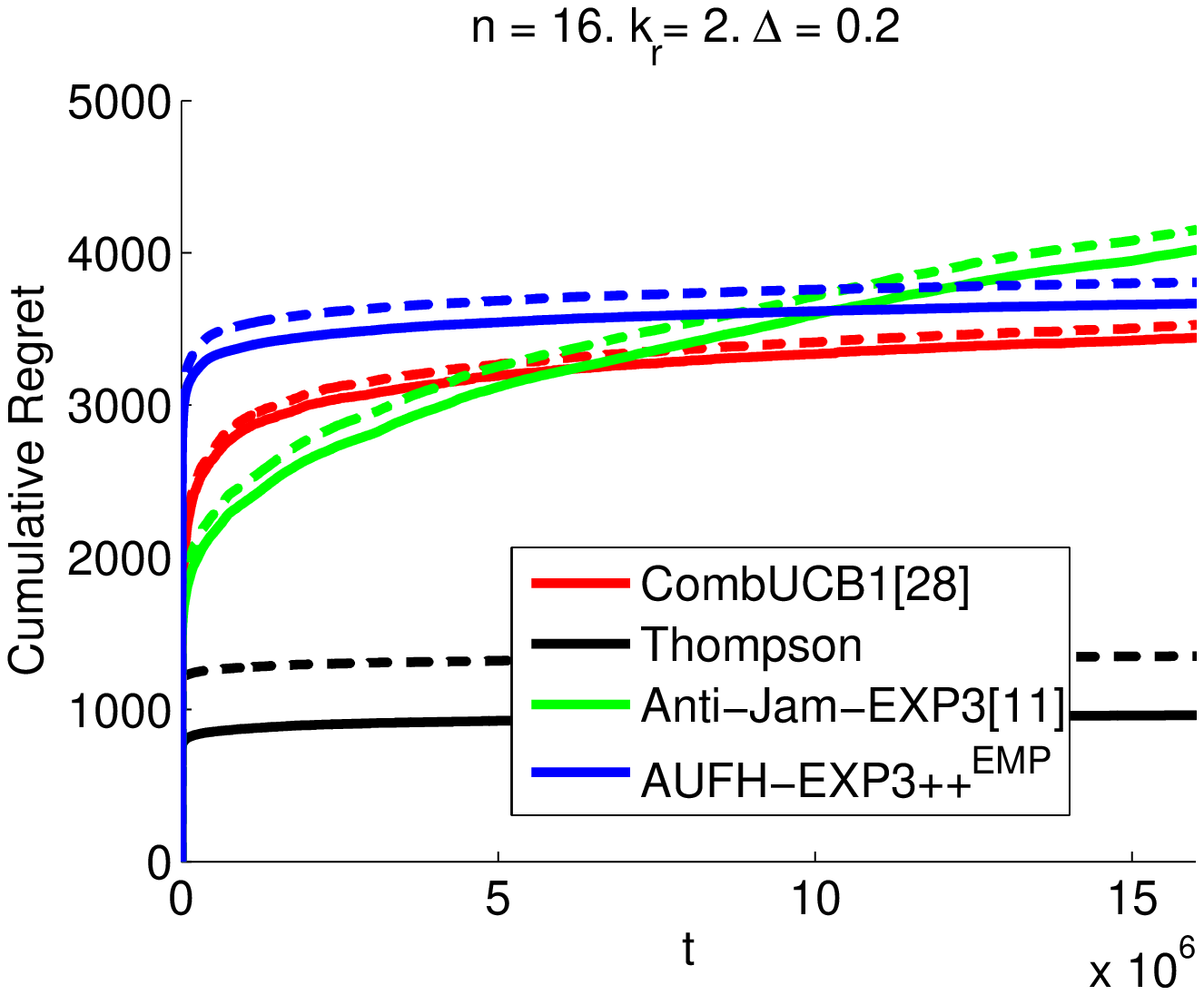}
 \caption{Performance Comparison in the Contaminated Stochastic Regime. }
\end{figure*}

\begin{figure*}
\includegraphics[width=2.33in]{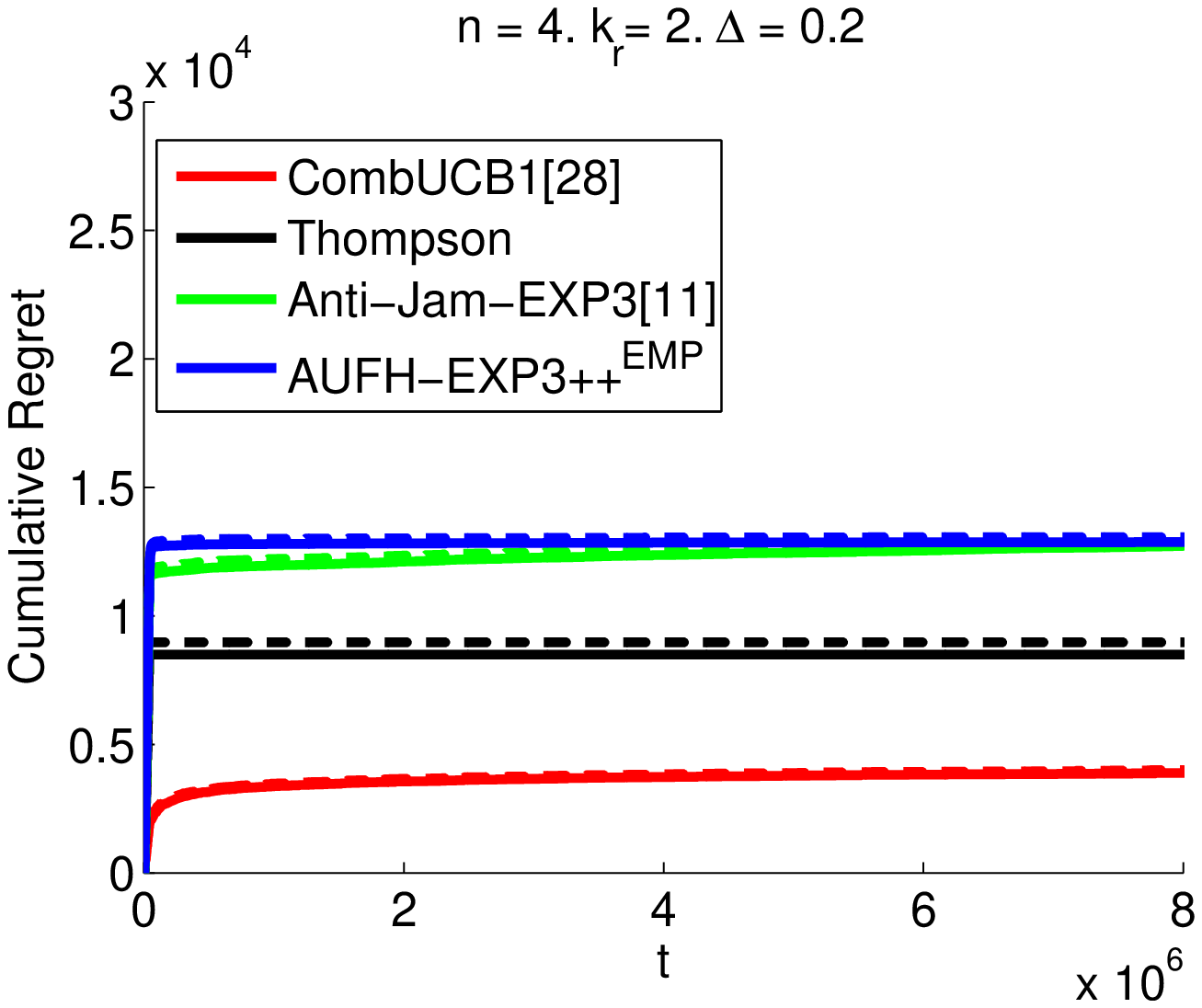}
\includegraphics[width=2.33in]{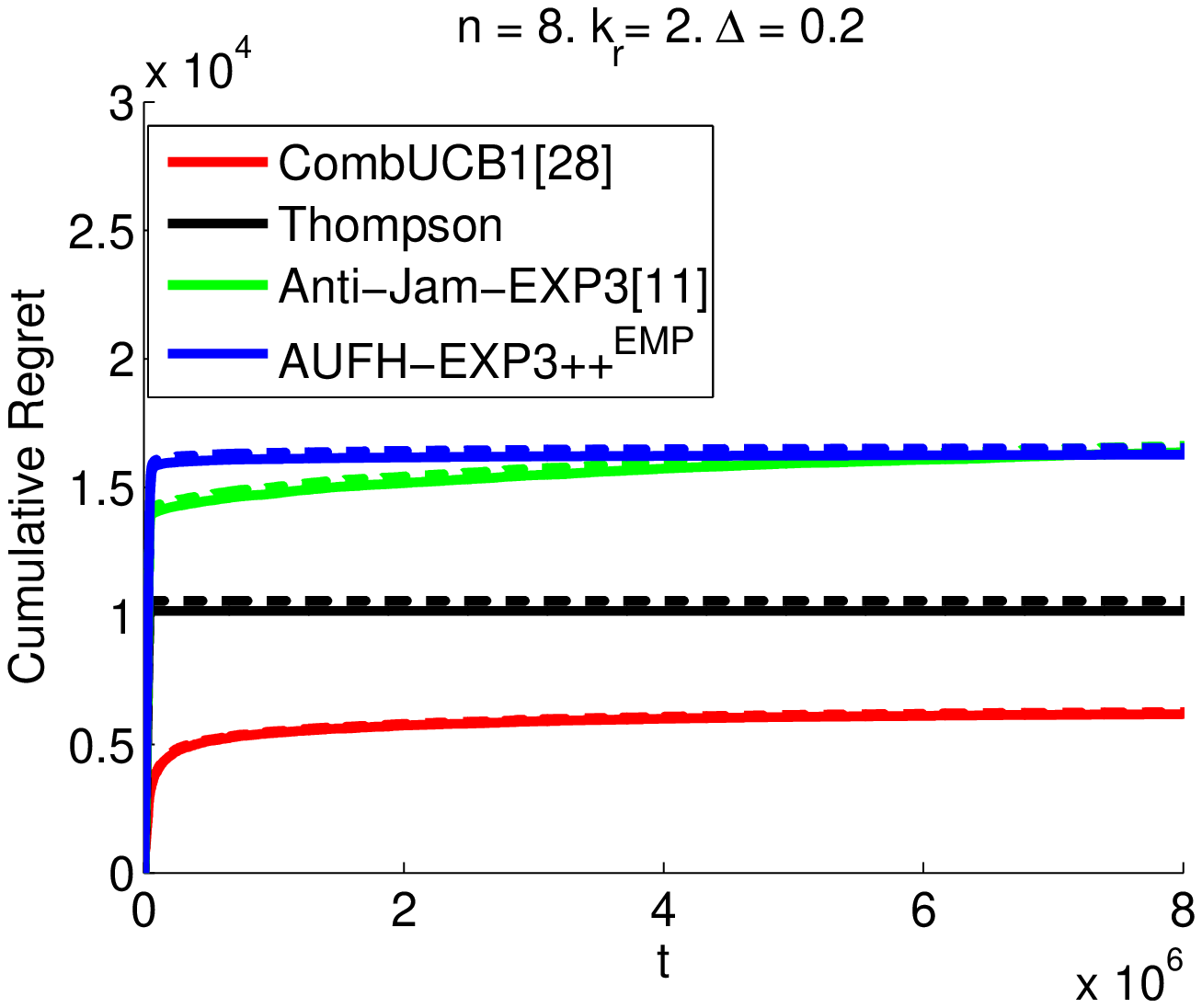}
\includegraphics[width=2.33in]{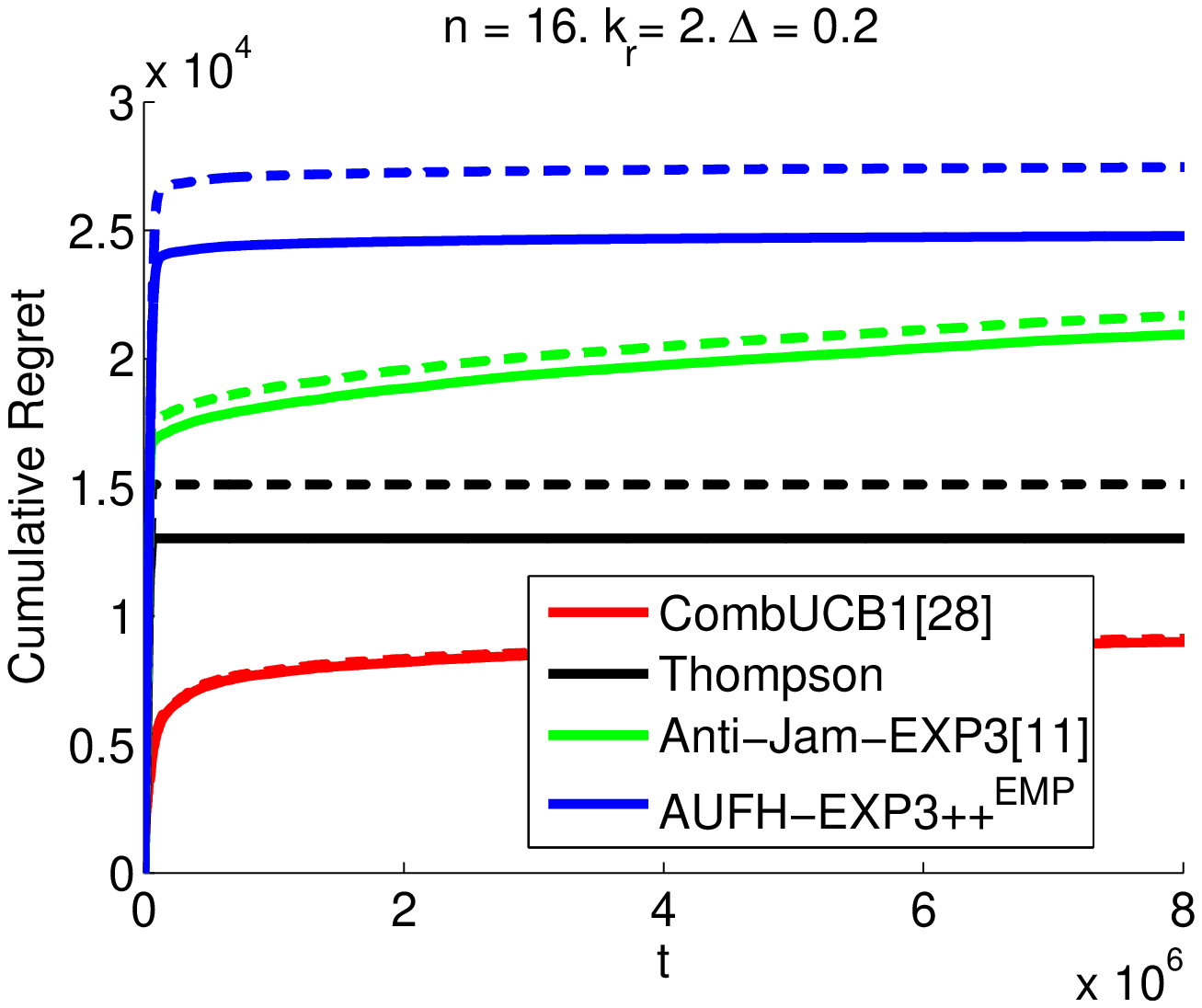}
  \caption{Performance Comparison in the Oblivious Adversarial Regime.}
\end{figure*}



\begin{figure*}
\vspace{-.4cm}
\includegraphics[width=2.33in]{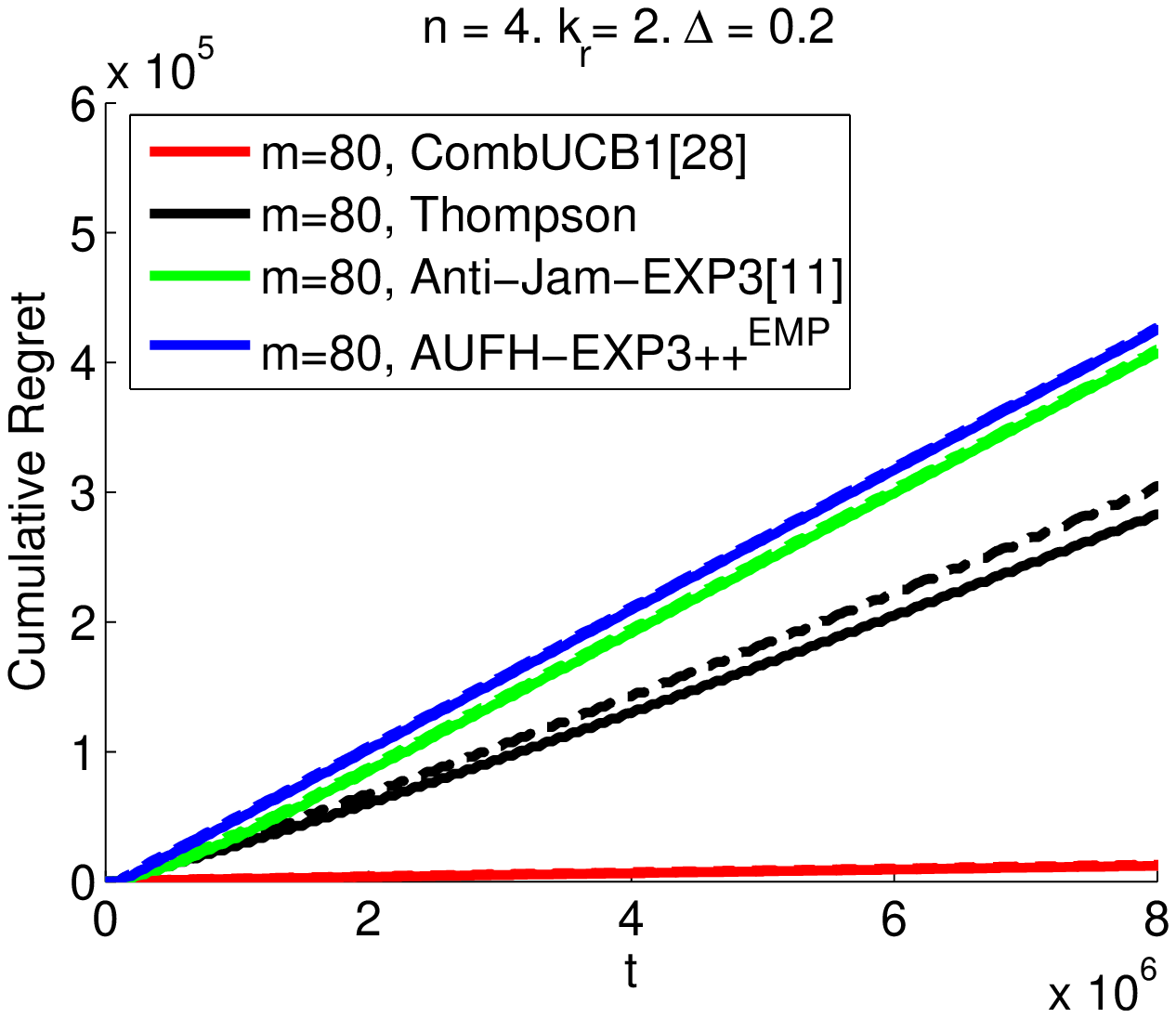}
\includegraphics[width=2.33in]{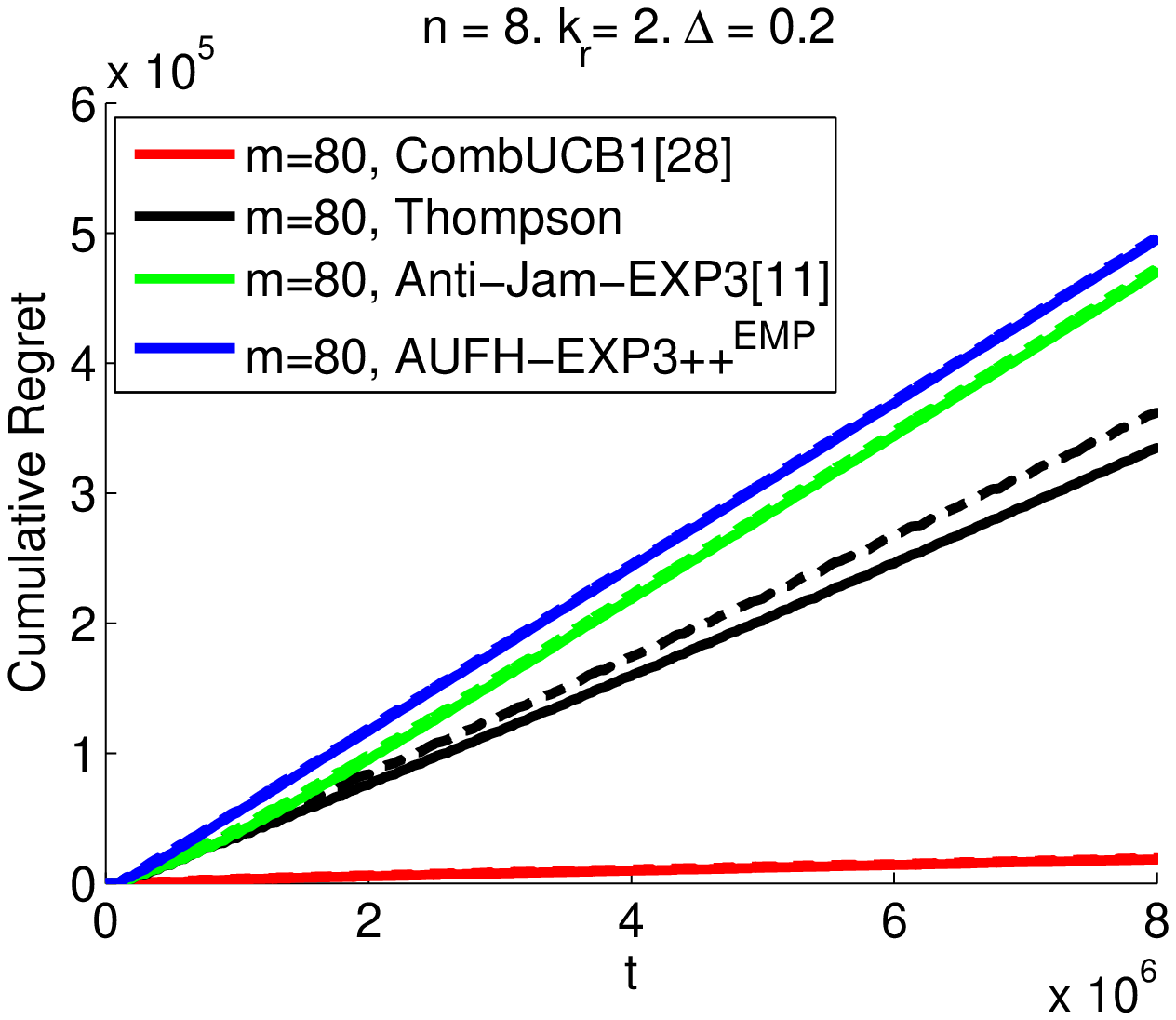}
\includegraphics[width=2.33in]{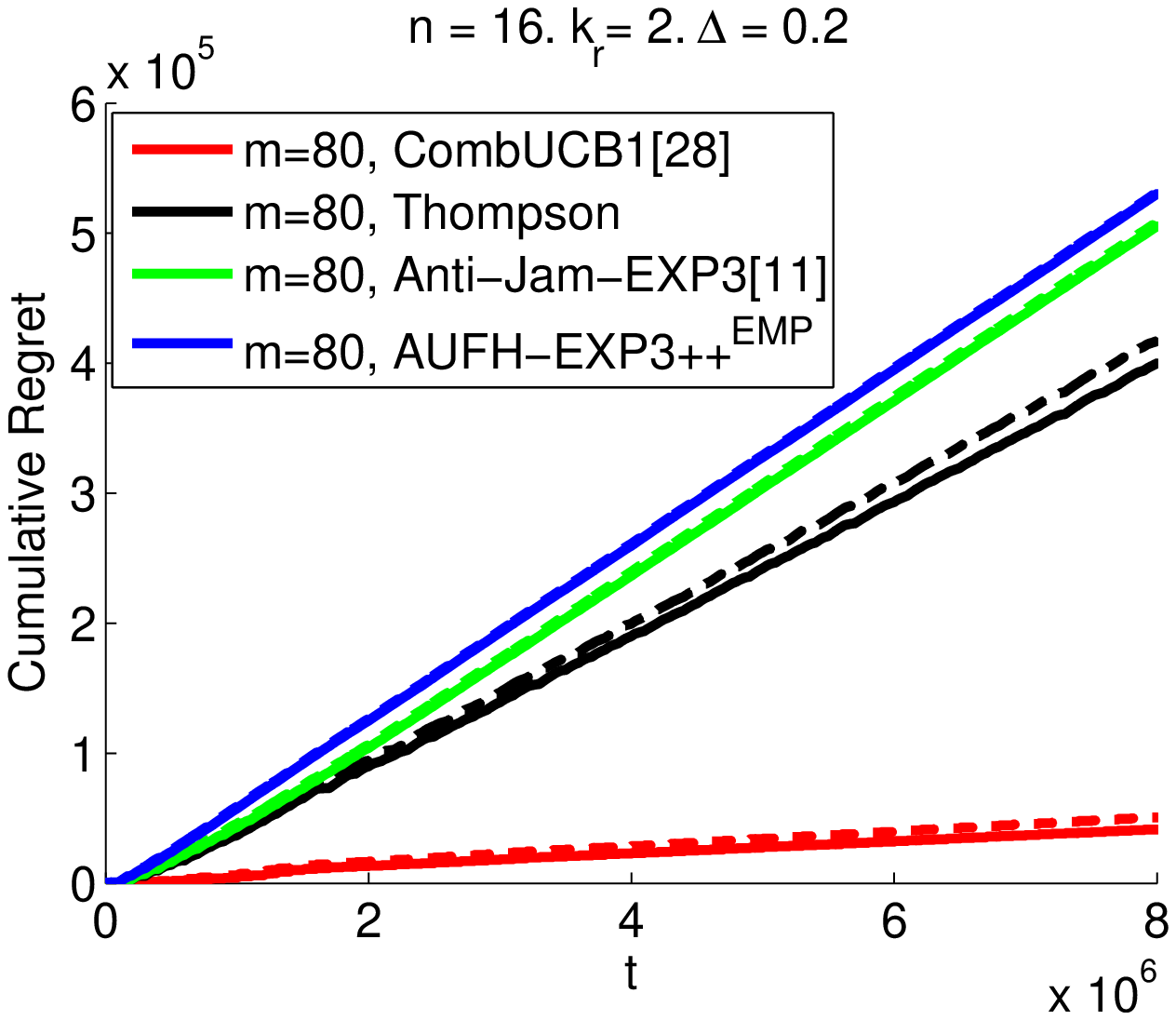}
  \caption{Performance Comparison in the Adaptive Adversarial  Regime.}
\end{figure*}

\section{Implementation Issues and Simulation Results}
%
In this section, we consider the wireless communications from a transmitter to a receiver  that is by default in the stochastic regime with Bernoulli distributions for rewards.
W.l.o.g., we assume a constant unitary data packet rate from the
transmitter for each channel $k_t \subseteq S_t$ over every time slot $t$, i.e. $M=1$ packet, where $k_t=4$. All experiments were conducted on an
 off-the-shelf desktop with dual
$6$-core Intel i7 CPUs clocked at $2.66$Ghz. For all the suboptimal channels the rewards are Bernoulli with bias $0.5$, and we set a single
best channel whose reward is Bernoulli with bias $0.5+\Delta$.

 To show the advantages of our AUFH-EXP3++ algorithms, we
compare their performance  to other existing MAB based algorithms, which includes:
the EXP3 based anti-jamming algorithm in \cite{QianJSAC12}, and we named it as ``Anti-Jam-EXP3"; The combinatorial UCB-based algorithm
``CombUCB1" with almost tight regret bound as proved in \cite{Branislav2015}; the combinatorial version of the Thompson's sampling algorithm \cite{Thomp33}. Here we consider the use of the Thompson's sampling algorithm for comparison due to its empirically good performance indicated in \cite{Seldin13}.
We make ten repetitions of each experiment to reduce the performance bias. In Fig. 2-5, the solid lines in the graphs  represent the mean performance over the experiments and the dashed lines represent the mean plus on standard deviation (std) over the ten repetitions of the corresponding experiments. Note that, for a given optimal channel access strategy, small regret values indicate the large number of data packets reception.

At first,  we run our experiments by choosing different size of available channels
 $n=8, 16, 60$. The size of receiving
channels and gap is always $k_r=4$ and $\Delta = 0.2$, respectively.  Our first set of experiments shown in Fig. 2, we run each of the algorithm for $10^7$ rounds. We choose $(n,k_r)$ pairs
equals to $(8,4), (16,4), (60,4)$ to see how our algorithms perform from a small size of channel access strategy set ($\binom{8}{4}=70$) to a large size of  channel access strategy set ($\binom{60}{4}=487635$).  For different versions of our AUFH-EXP3++ algorithms, they are
parameterized by ${\xi _t}(f) = \frac{{\ln (t{{\hat \Delta }_t}{{(f)}^2})}}{{32t{{\hat \Delta }_t}{{(f)}^2}}}$, where ${{{\hat \Delta }_t}(f)}$ is the
empirical estimate of ${{{ \Delta }_t}(f)}$ defined in (\ref{eq:EstD}). The target of our experiment is to demonstrate
 that in the stochastic regime the exploration parameters are in full control of the performance we run the AUFH-EXP3++ algorithm with two different
 learning rates. AUFH-EXP3++$^{EMP}$ corresponds to $\eta_t= \beta_t$ and AUFH-EXP3++$^{ACC}$ corresponds to $\eta_t= 1$. Note that only AUFH-EXP3++$^{EMP}$
 has a performance guarantee in the adversarial regime. For our AUFH-EXP3++ algorithms, we transform the rewards into losses via
 $\ell_t(f)= 1- g_t(f)$, other algorithms operate directly on the rewards.

 From the results presented in Fig. 2, we see that in all the experiments, the performance of AUFH-EXP3++$^{EMP}$ is almost
 identical to the performance of CombUCB1. That means our algorithm can attain almost optimal transmission efficiency  in stochastic environments, and
 our algorithm scales well in the large channel access strategy setting.
Thus,  AUFH-EXP3++$^{EMP}$  has all advantages of the stochastic MAB algorithms, and has much better performance gain than Anti-Jam-EXP3 \cite{QianJSAC12}. Moreover, unlike CombUCB1 and Thompson's sampling, AUFH-EXP3++$^{EMP}$ is secured against a potential adversary during the wireless communications game. In addition, the AUFH-EXP3++$^{ACC}$ algorithm  can be seen as a special teaser to show the algorithm performance in the condition of $\eta_t > \beta_t$. It performs better than  AUFH-EXP3++$^{EMP}$, but it does not have the adversarial regime performance guarantee.

\begin{figure*}
\centering
\includegraphics[scale=.42]{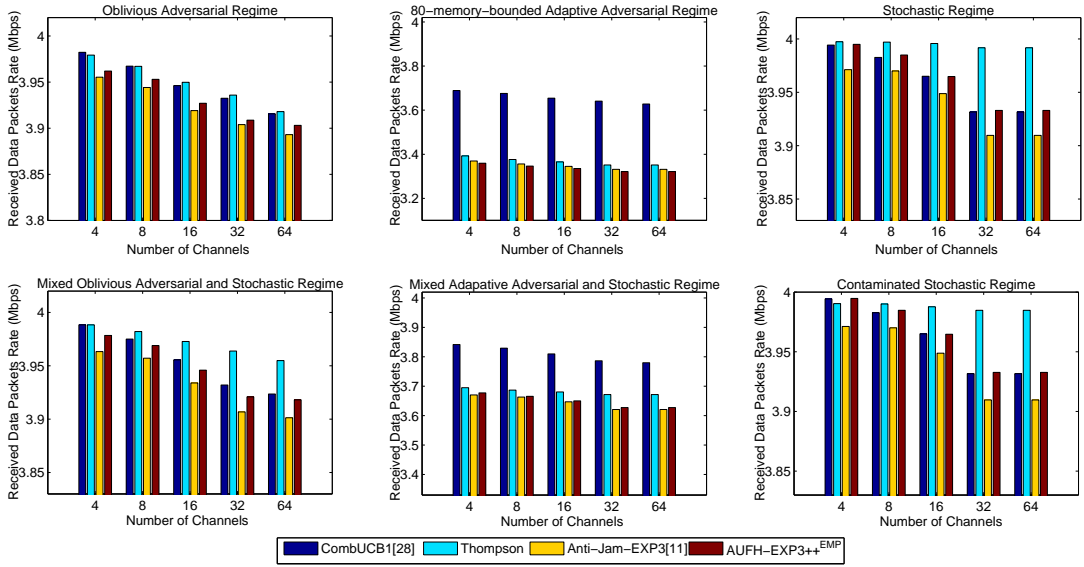}
\caption{Received Data Packets Rate in Different Regimes. The legend
below corresponds to all figures.}
\label{fig:digraph}
\vspace{.2cm}
\end{figure*}

In our second set of experiments, we simulate moderately contaminated stochastic environment
by drawing the first 2,500 rounds of the game according
to one stochastic model and then switching the best channel
and continuing the game until $8*10^6$ rounds. This action can be regarded as an occasional jamming behavior.
  In this case, the contamination is not fully adversarial, but drawn
from a different stochastic model.  We run this experiment
with $\Delta = 0.2 $, $k_r=2$  and $n=4, 8, 16$ to see the noticed leaning performance. The results are
presented in Fig. 3.  Although it is hard to see the first 2,500 rounds
on the plot, their effects on all the algorithms is clearly
visible. Despite the initial corrupted rounds the AUFH-EXP3++$^{EMP}$
algorithm successfully returns to the stochastic operation
mode and achieves better results than Anti-Jam-EXP3 \cite{QianJSAC12}.


To the best of our knowledge, it is very hard to simulate the fully adversarial regime with arbitrarily changing
oblivious jammer. In our third set of experiments shown in Fig. 4, we emulate the
adversary regime under oblivious jamming attack by setting the $\Delta$ value of  the best channel randomly
from $[0.1, 0.3]$ and switch  the best
channel to different indices of channels in the channel set at every other time slot by a pseudorandom sequence generator function. The channel rewards are
 determined before running the algorithm. It is not difficult to feel that the reward sequences still
follow certain stochastic pattern, but not that obvious. We set the typical parameter $k_r=2$,  $\Delta= 0.2$ and run all the algorithms up to $8*10^6$ rounds. It can be found that our AUFH-EXP3++$^{EMP}$
algorithm will be close to and have slightly better performance when compared to Anti-Jam-EXP3 \cite{QianJSAC12}, which confirms with our theoretical
analysis. 


In our fourth set of experiments shown in Fig. 5, we simulate the adaptive jamming attack case in the adversarial regime with a typical memory  $m=80$. We can see large performance degradations for all algorithms when compared to the oblivious jammer case. We can find that the performance of AUFH-EXP3++$^{EMP}$ and Anti-Jam-EXP3 \cite{QianJSAC12}  still enjoys the almost the same regret performance, and their large regrets
indicate their sensitiveness to the adaptive jammer.

\begin{table*}\centering
\caption{Computation Time Comparisons of Algorithm 1 and Algorithm 2}
\begin{tabular}{|l|l|l|l|l|l|l|l|}
\hline

\multicolumn{8}{|c|}{$(n, k_r)$} \\

\hline
 Alg. Ver. \emph{vs} Comp. Time (micro seconds)  
  & $(12, 4)$ & $(24, 4)$ & $(48, 6)$ & $(48, 12)$ & $(64, 6)$  & $(64, 12)$ & $(64, 24)$ \\
\hline
AUFH-EXP3++$^{EMP}$:¡¡Algorithm1          & 23 & 167 & 699 & 2247 & 8375 & 162372 & 862961  \\
AUFH-EXP3++$^{EMP}$:¡¡Algorithm2          & 4 & 9 & 31 & 57 &  74 & 134 &  280  \\
\hline
\end{tabular}

\end{table*}

We also compared the computing time of the two versions of AUFH-EXP3++$^{EMP}$, Algorithm 1 and Algorithm 2, with different set of $(n,k_r)$
 pairs for each round. The results are listed in table I. From the results, we can see that Algorithm 2 scales linearly with the
 increase of the size of $n$ and $k_r$, and have very low computational cost than the Algorithm 1. Imagine in a practical typical multi-channel
 wireless communication system with $(n, k_r)= (64, 12)$, the Algorithm 1 takes about $162$ seconds to finish one round calculation that is
 infeasible, while the Algorithm 2 takes about $.134$ seconds to finish one round calculation that is very reasonable in practical implementation.


For brevity, we do not plot the regret performance figures for the mixed adversarial and stochastic regime. However, in our last experiments, we
compare the received data packets rate (Mbps) for all the four different regimes after a relative long period of learning rounds $t=2*10^7$. Here we assume $ M=1$ packet contains $1000$ bits and
each time slot is just one second. We set $k_r=2$ and $\Delta= 0.2$ as fixed values for all different size of channel set $n$. We plot our results
in Fig. 6. It is easy to find that our algorithm AUFH-EXP3++$^{EMP}$ attains almost all the advantages of the stochastic MAB algorithms CombUCB1, and
has better throughput performance than Anti-Jam-EXP3. As we have noticed, we also put the results of CombUCB1 \cite{Branislav2015} in the oblivious adversarial, adaptive adversarial and
contaminated regimes, etc., although the algorithm is not applicable in theory. This proves that our proposed algorithm AUFH-EXP3++ can be
applied for general unknown communication environments in different regimes with flexibility. Interestingly, we find that the Thompson's sampling algorithm \cite{Thomp33} performs superiorly in all regimes, and this empirical fact is observed in the machine learning society. We believe it is a promising direction to study  its theoretical ground from the beginning for the  collected (security) non--i.i.d. data inputs.

\section{Conclusion and Future Works}
In this paper, we have proposed the first adaptive multichannel-access algorithm for wireless communications without the knowledge about the nature of environments. At first, we captured the feature of the general wireless environments and divided them into four regimes, and then provided solid theoretical analysis for each of them. Through theoretical analysis, we found that the
almost optimal performance is achievable for all regimes.  Extensive simulations were conducted  to verify the learning performance of our algorithm in different regimes and much better performance
improvements over classic approaches. The proposed algorithm could be implemented efficiently in practical wireless communication systems with different sizes.
 Our framework is of general value, which can be extended by incorporating power control module based on estimated gradient algorithms (under bandit feedback), taking power budgets into account and accessing problems based on observed side information (as ``contextual bandit" \cite{Bubeck12}) for wireless communication scenarios under unexpected security attacks. The idea of this work could also be combined with
 other online learning-based channel prediction algorithms to perform the  joint optimal resource allocation with the configuration of physical layer techniques, such as the MIMO channel and its power allocations.
  We plan to extend our proposed  algorithms to general combinatorial settings and forecast that their variants can be applied in many practical tough environments for wireless networks monitoring, secure routing problems, rumors propagation in social networks (with contextual bandit setting), etc.



 \begin{IEEEbiography}[{\includegraphics[width=2.6in,height=1.25in,clip,keepaspectratio]{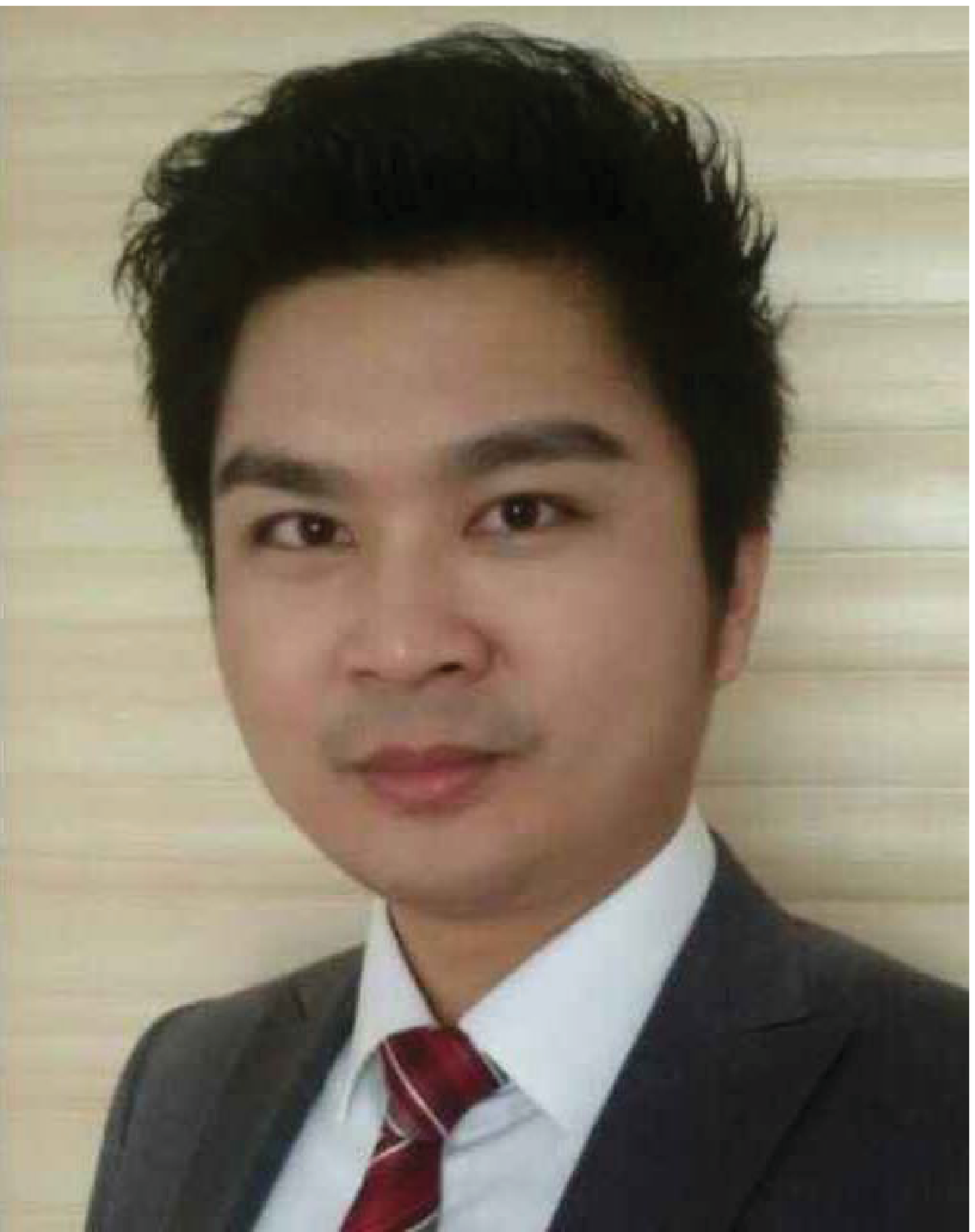}}]{Pan Zhou(S'07--M'14)} is currently an associate professor
 with School of Electronic Information and Communications, Huazhong University of Science and Technology, Wuhan, P.R. China. He received his Ph.D. in the School of Electrical and Computer Engineering at the Georgia
Institute of Technology (Georgia Tech) in 2011, Atlanta, USA. He
received his B.S. degree in the \emph{Advanced Class} of
HUST, and a M.S. degree in the Department of Electronics and Information Engineering
from HUST, Wuhan, China, in 2006 and 2008, respectively.
He held honorary degree in his bachelor and merit research award
of HUST in his master study. He was a
senior technical memeber at Oracle Inc, America during 2011 to 2013, Boston, MA, USA,  and worked on hadoop and distributed storage system for big data
analytics at Oralce cloud Platform.  His current research interest includes:  communication and information networks, security and privacy,  machine learning and big data.
\end{IEEEbiography}

 \begin{IEEEbiography}[{\includegraphics[width=2.6in,height=1.25in,clip,keepaspectratio]{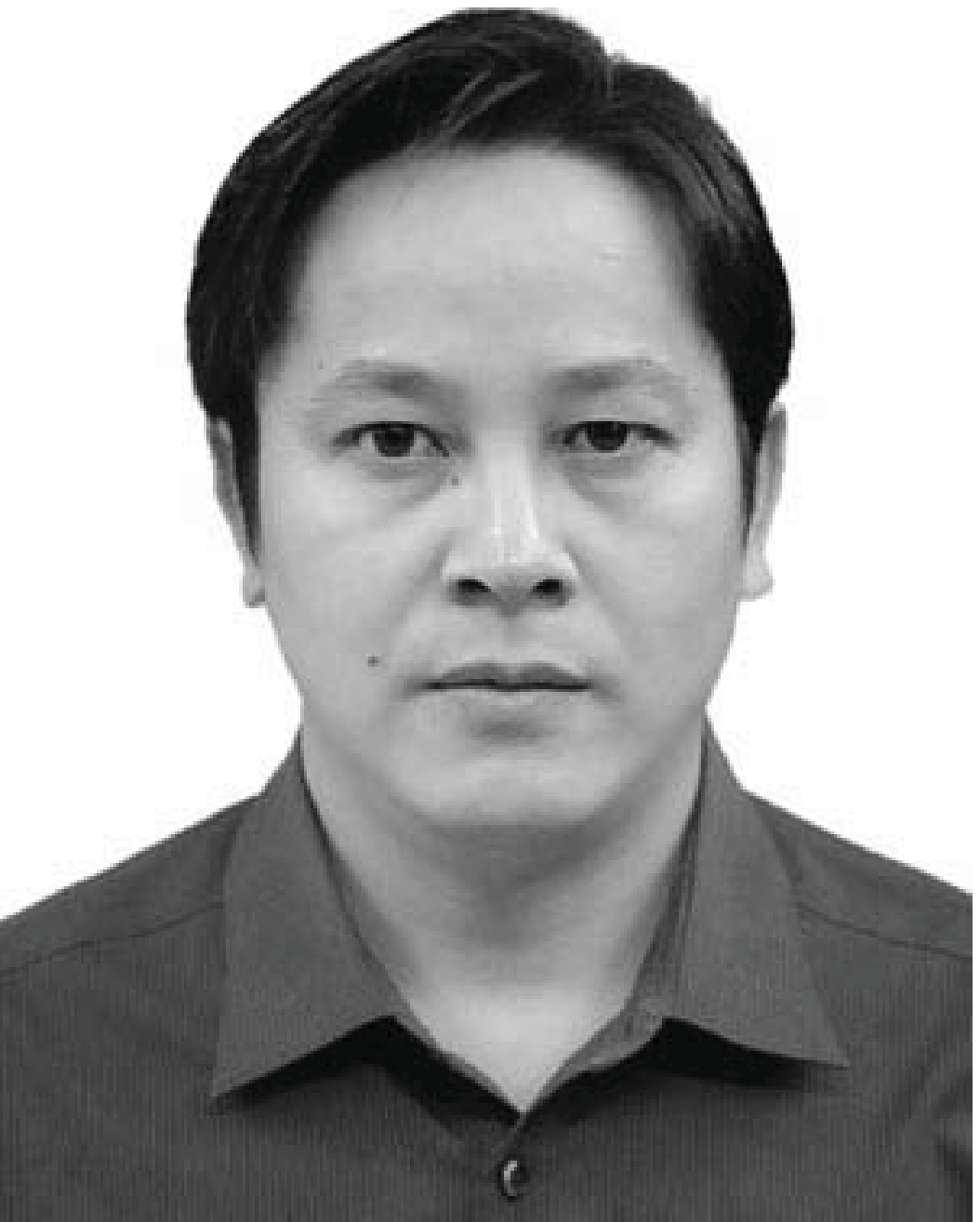}}]{Tao Jiang (M'06--SM'10)}
is currently a Distinguished Professor in the School of Electronics Information and Communications, Huazhong University of Science and Technology, Wuhan, P. R. China. He received the B.S. and M.S. degrees in applied geophysics from China University of Geosciences, Wuhan, P. R. China, in 1997 and 2000, respectively, and the Ph.D. degree in information and communication engineering from Huazhong University of Science and Technology, Wuhan, P. R. China, in April 2004. From Aug. 2004 to Dec. 2007, he worked in some universities, such as Brunel University and University of Michigan-Dearborn, respectively. He has authored or co-authored over 200 technical papers in major journals and conferences and 8 books/chapters in the areas of communications and networks. He served or is serving as symposium technical program committee membership of some major IEEE conferences, including INFOCOM, GLOBECOM, and ICC, etc.. He is invited to serve as TPC Symposium Chair for the IEEE GLOBECOM 2013, IEEEE WCNC 2013 and ICCC 2013. He is served or serving as associate editor of some technical journals in communications, including in IEEE Transactions on Signal Processing, IEEE Communications Surveys and Tutorials, IEEE Transactions on Vehicular Technology, and IEEE Internet of Things Journal, etc.. He is a recipient of the NSFC for Distinguished Young Scholars Award in 2013, and he is also a recipient of the Young and Middle-Aged Leading Scientists, Engineers and Innovators by the Ministry of Science and Technology of China in 2014. He was awarded as the Most Cited Chinese Researchers in Computer Science announced by Elsevier in 2014. He is a senior member of IEEE.
\end{IEEEbiography}

\end{document}